\documentclass[sigplan,10pt,nonacm]{acmart}
\renewcommand\footnotetextcopyrightpermission[1]{}
\graphicspath{{figures/}{results/}}

\usepackage{latexsym}
\usepackage[export]{adjustbox}
\usepackage{hyperref}
\usepackage{subfigmat}
\usepackage{multirow}
\usepackage{tikz}
\usepackage{array}
\usepackage{xspace}
\usepackage{multirow}
\usepackage{comment}
\usepackage{url}
\usepackage{xurl}
\usepackage[normalem]{ulem}
\usepackage{mdwlist}
\usepackage{algorithm}
\usepackage{amsmath,amsthm}
\usepackage{booktabs}
\usepackage[noend]{algpseudocode}
\usepackage{centernot}
\usepackage{fancyvrb}
\usepackage{listings}
\usepackage{alltt}
\usepackage{graphicx}
\usepackage{placeins}
\usepackage{caption}
\usepackage[ampersand]{easylist}
\usepackage{fancybox}
\usepackage{tcolorbox}
\usepackage{balance}
\usepackage{xcolor}
\usepackage{natbib}
\usepackage{makecell}
\ifPDFTeX
  \usepackage{flexisym}
\fi
\usepackage{pifont}
\usepackage{mathtools}
\usepackage{enumitem}
\usepackage{arydshln}

\usepackage[ampersand]{easylist}
\usepackage{fancybox}
\usepackage{verbatim}

\usepackage[framemethod=tikz]{mdframed}
\usepackage{amsfonts}

\usepackage{multirow}

\usepackage{pifont}

\usepackage{cleveref}
\crefname{section}{§}{§}
\Crefname{section}{§}{§}

\crefformat{section}{§#2#1#3}
\Crefformat{section}{§#2#1#3}

\definecolor{electricyellow}{rgb}{1.0, 1.0, 0.0}

\algnewcommand\algorithmicforeach{\textbf{for each}}
\algdef{S}[FOR]{ForEach}[1]{\algorithmicforeach\ #1\ \algorithmicdo}

\definecolor{light-gray}{gray}{0.8}
\definecolor{amethyst}{rgb}{0.6, 0.4, 0.8}
\definecolor{ForestGreen}{RGB}{34,139,34}

\makeatletter
\newcommand{\smallsym}[2]{#1{\mathpalette\make@small@sym{#2}}}
\newcommand{\make@small@sym}[2]{%
  \vcenter{\hbox{$\m@th\downgrade@style#1#2$}}%
}
\newcommand{\downgrade@style}[1]{%
  \ifx#1\displaystyle\scriptstyle\else
    \ifx#1\textstyle\scriptstyle\else
      \scriptscriptstyle
  \fi\fi
}
\makeatother

\newcommand\pagebudget[1]{}
{}
{}

{}

\global\let\shownotes\empty
\ifx \shownotes \empty

\newcommand\missing[1]{\textcolor{red}{#1}}

\newcommand\removed[1]{\textcolor{red}{\sout{#1}}}

\else

\newcommand\missing[1]{}

\newcommand\removed[1]{}

\fi

\newcommand\OMIT[1]{}

\definecolor{orange}{rgb}{0.9,0.4,0.0}
\definecolor{purple}{rgb}{0.8,0,0.7}

\definecolor{codeblue}{rgb}{0,0,1}
\definecolor{codegreen}{rgb}{0,0.6,0}
\definecolor{codegray}{rgb}{0.5,0.5,0.5}
\definecolor{codepurple}{rgb}{0.58,0,0.82}
\definecolor{backcolour}{rgb}{0.95,0.95,0.92}
\definecolor{nocolor}{rgb}{1,1,1}
\definecolor{red}{rgb}{0.6,0,0} 
\definecolor{blue}{rgb}{0,0,0.6}
\definecolor{green}{rgb}{0,0.8,0}
\definecolor{cyan}{rgb}{0.0,0.6,0.6}
\definecolor{lightgray}{gray}{0.98}
\definecolor{lightblue}{rgb}{0.13, 0.67, 0.8}
\definecolor{lightorange}{RGB}{255,247,230}
\definecolor{codegreen}{rgb}{0,0.6,0}
\definecolor{codegray}{rgb}{0.5,0.5,0.5}
\definecolor{codepurple}{rgb}{0.58,0,0.82}
\definecolor{keywordcolor}{RGB}{94,20,64}
\definecolor{bluekeywords}{rgb}{0,0,1}
\definecolor{greencomments}{rgb}{0,0.5,0}
\definecolor{redstrings}{rgb}{0.64,0.08,0.08}
\definecolor{xmlcomments}{rgb}{0.5,0.5,0.5}
\definecolor{types}{rgb}{0.17,0.57,0.68}
\definecolor{KWColor}{RGB}{0,0,255}
\definecolor{AnnotationColor}{RGB}{0,137,180}
\definecolor{BlackColor}{RGB}{0,0,0}
\definecolor{CommentColor}{rgb}{0.12,0.38,0.18}
\definecolor{StringColor}{rgb}{0.06,0.10,0.98}
\definecolor{darkred}{rgb}{0.65,0,0}
\definecolor{lightgrey}{rgb}{0.8,0.8,0.8}
\definecolor{marmalade}{RGB}{193,101,18}
\definecolor{peach}{RGB}{250,217,193}
\definecolor{lime}{RGB}{220,237,193}

\definecolor{highlight}{RGB}{255,255,150}
\definecolor{lightgray}{RGB}{240,240,240}
\definecolor{white}{RGB}{255,255,255}
\definecolor{tacao}{RGB}{234, 182, 118}
\definecolor{softblue}{RGB}{118, 170, 234}

\ifdefined\NewTColorBox
  \NewTColorBox{NewBox}{ s O{!htbp} }{%
    floatplacement={#2},
    IfBooleanTF={#1}{float*,width=\textwidth}{float},
    colframe=blue!50!black,colback=blue!10!white,
  }
\fi
  
\tcbset{
    outerbox/.style={
        colframe=black, 
        colback=white, 
        arc=2mm, 
        boxrule=1pt, 
        left=2mm,
        right=2mm,
        top=2mm,
        bottom=2mm,
        boxsep=2mm,
        toptitle=1mm,
        bottomtitle=1mm,
        colbacktitle=white,
        coltitle=black,
        fonttitle=\bfseries,
    },
    innerbox/.style={
        colback=tacao, 
        arc=0mm, 
        boxrule=1pt, 
        boxsep=0mm, 
        left=2mm,
        right=2mm,
        top=2mm,
        bottom=2mm,
    },
    innerbox2/.style={
        colback=softblue, 
        arc=0mm, 
        boxrule=1pt, 
        boxsep=0mm, 
        left=2mm,
        right=2mm,
        top=2mm,
        bottom=2mm,
    }
}

\mdfdefinestyle{background}{backgroundcolor=lightorange,innerrightmargin=0cm,innertopmargin=-0.1cm,innerbottommargin=-0.10cm,leftmargin=+0cm, roundcorner=2pt}

\newif\ifanon\anonfalse

\ifanon

\else

\fi

\usepackage[textsize=tiny,textwidth=0.6in]{todonotes}
\newcommand{\system}{\textsc{EdgePace}\xspace}

\setcopyright{none}
\copyrightyear{2026}
\acmYear{2026}
\acmDOI{}

\begin{document}

\title{\system: Deadline- and Freshness-Aware LLM Scheduling on Unified-Memory Platforms}

\author{Teng Mei}
\authornote{Teng Mei and Chengxuan Pei contributed equally to this work.}
\affiliation{%
  \institution{The Chinese University of Hong Kong, Shenzhen}
  \city{Shenzhen}
  \country{China}
}
\email{mei347957756@gmail.com}

\author{Chengxuan Pei}
\authornotemark[1]
\affiliation{%
  \institution{The Chinese University of Hong Kong, Shenzhen}
  \city{Shenzhen}
  \country{China}
}
\email{chengxuanpei@link.cuhk.edu.cn}

\author{Weijun Wang}
\affiliation{%
  \institution{Institute for AI Industry Research (AIR), Tsinghua University}
  \city{Beijing}
  \country{China}
}
\email{wangweijun@air.tsinghua.edu.cn}

\author{Jie Wu}
\affiliation{%
  \institution{Fudan University}
  \city{Shanghai}
  \country{China}
}
\email{jwu@fudan.edu.cn}

\author{Marco Canini}
\affiliation{%
  \institution{KAUST}
  \city{Thuwal}
  \country{Saudi Arabia}
}
\email{marco@kaust.edu.sa}

\author{Yunming Xiao}
\authornote{Corresponding author.}
\affiliation{%
  \institution{The Chinese University of Hong Kong, Shenzhen}
  \city{Shenzhen}
  \country{China}
}
\email{yunmingxiao@cuhk.edu.cn}

\renewcommand{\shortauthors}{Teng Mei et al.}

\begin{abstract}

Large language models (LLMs) increasingly run on edge devices alongside tasks that continuously process sensor data. In many applications, these foreground tasks prepare observations as inputs to the LLM, whose outputs guide subsequent processing or actions. The observations must be processed by foreground deadlines, and LLM feedback must arrive while its source observations remain useful. Yet the CPU and accelerators share a unified-memory system, so LLM inference can delay foreground processing through memory-bandwidth contention even when compute resources remain available. We present \system, a bandwidth-aware runtime that protects foreground deadlines while sustaining timely LLM feedback. \system creates safe suspension points across prefill and decode, monitors foreground memory demand and stalls at low cost, and schedules inference according to contention risk, deadline slack, and output freshness. We evaluate \system on a OnePlus 13 and three NVIDIA Jetson platforms using multi-camera processing and robotic target tracking. Relative to unmanaged inference, it reduces foreground deadline misses by 24.1\%--96.1\% while retaining 64.9\%--98.5\% of LLM throughput. All measured LLM outputs arrive within their task-defined freshness windows. On OnePlus 13, for example, camera deadline misses fall by 91.3\% while 90.4\% of LLM throughput is retained.
  
\end{abstract}




\maketitle

\section{Introduction}

Large language models (LLMs) have been increasingly pushed to resource-constrained edge devices~\cite{xu2025llmnpu,yi2025edgemoe,wang2025d2moe, yin2025elastilm}, e.g., smartphones and NVIDIA Jetson boards, to enable interactive assistance and screen-based task automation~\cite{wen2024autodroid}, as well as speech understanding and perception of the local environment~\cite{yuan2024mobilefm}. In these applications, LLM inference is part of a broader workflow: the device processes sensor data to inform the LLM while also running perception and control tasks~\cite{wang2025agile3d,jung2025aria}. Running the LLM locally lets this workflow respond without a cloud round-trip and keeps sensitive sensor data on the device.

Despite these benefits, LLM inference consumes a substantial share of an edge device's resources. In many applications, the LLM and foreground sensor-processing tasks run concurrently on the same device and share compute and memory resources (e.g., due to power, thermal or budget constraints). Foreground tasks face tight timing budgets, as delays can cause missed deadlines, dropped frames, or wrong decisions based on stale observations~\cite{han2024pantheon,xin2026sereno}. The system must therefore sustain useful LLM performance while preserving the timely flow of sensor data through the workflow.



\begin{figure}
    \centering
    \subfigure[p95 latency]{%
        \includegraphics[width=0.49\linewidth, trim=0mm 5mm 0mm 0mm]{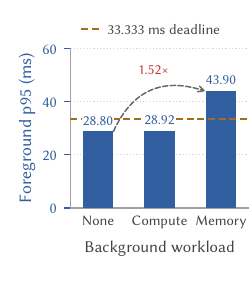}%
        \label{fig:comparison-latency}}\hfill
    \subfigure[Deadline misses]{%
        \includegraphics[width=0.49\linewidth, trim=0mm 5mm 0mm 0mm]{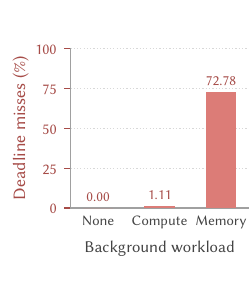}%
        \label{fig:comparison-misses}}
    \vspace{-0.1in}
    \caption{Compared to the compute-intensive background workload, the memory-intensive workload (a) delays the foreground task by $1.52\times$ on unified-memory devices and (b) causes 72.78\% deadline misses.
    }
    \Description{Two side-by-side bar charts compare three background
    conditions: none, GPU compute, and GPU memory traffic. Left: foreground
    p95 latencies are 28.80, 28.92, and 43.90 ms. Right: deadline-miss rates
    are 0, 1.11, and 72.78 percent. The latency panel includes a 33.333 ms
    deadline line and a single arrow from the baseline bar to the GPU
    memory bar, annotated with their 1.52 times latency ratio.}
    \label{comparison}
\end{figure}

To understand the resource contention, we conduct a thorough analysis and identify the shared memory subsystem as the source. 
Edge devices such as smartphones and Jetson boards typically share a \emph{unified} memory among all compute
units: the CPU, the GPU, and NPU~\cite{xu2025llmnpu,jetson}. Unified memory simplifies data sharing between these units, but their concurrent memory traffic must share limited memory bandwidth. 
Model-weight transfers and KV-cache accesses make LLM inference a substantial consumer of that bandwidth~\cite{yi2025edgemoe,zhu2025nanoflow,xin2026sereno}. 
Meanwhile, sensor-processing tasks on CPU must also use the same memory system while needing to satisfy hard deadlines; so contention can delay their requests and cause deadline misses even when compute resources are not saturated~\cite{wang2025agile3d,han2024pantheon}. 

To isolate this effect, we have deployed a four-camera input workload on a Jetson Nano Orion Super platform~\cite{Jetson-Nano-Orion} alongside either no background task, a compute intensive GPU task, or a task generating sustained memory traffic. 
Figure~\ref{comparison} shows that the compute-intensive task has negligible interference with the four-camera task. In contrast, the sustained memory traffic increases the p95 latency of camera workload by 1.52$\times$ and raises the failure rate (i.e., deadline misses in the figure) to 72.78\%. 
Further, the interference also occurs with a real LLM background:
co-running a Qwen2.5-1.5B model~\cite{qwen2024qwen25} alongside a four-camera task will increase the deadline miss rate of the latter to 18.33\% (\cref{sec:eval-case1}).
These results motivate a runtime that regulates LLM execution according to foreground timing and shared-memory bandwidth pressure.
However, this is not straightforward. First, the runtime needs safe, frequent opportunities to pause LLM inference during both prefill and decode phases, without costly resumption or extensive changes to the LLM framework. Next, aggregate bandwidth alone may not be sufficient to indicate when foreground tasks are being delayed; the runtime needs more informative memory-pressure signals, but collecting them must not add appreciable interference. Moreover, scheduling must respond to dynamic scenarios with changing contention and foreground deadlines while allowing the LLM to finish before its input becomes stale.

We tackle the above problems with \system, a bandwidth-aware runtime that co-schedules LLM inference and foreground tasks on edge platforms. 
\system has three design pieces: $(i)$ a \emph{fine-grained execution plan} that partitions both prefill and decode phases at safe sublayer boundaries, allowing it to pause inference promptly and resume without repeating completed work; $(ii)$ a \emph{lightweight monitor} that samples hardware counters on foreground CPU threads and combines memory-demand and stall signals to detect pressure on foreground processing without burdening each scheduling decision; and $(iii)$ a \emph{scheduler} that uses these signals together with foreground deadline slack and the remaining freshness window of the LLM input to decide when inference should proceed or yield. Together, these components protect time-sensitive processing while allowing the LLM to make progress whenever resources permit.
We evaluate \system across four heterogeneous platforms, including three Jetson platforms~\cite{Jetson-Nano-Orion, nvidia2026jetsonThor, jetson} and a OnePlus 13 with Snapdragon 8 Elite~\cite{oneplus13specs}, and two workflows of multi-camera processing and robotic target tracking. Relative to unmanaged inference, \system reduces the foreground deadline miss rate by 24.1\%--96.1\% while retaining 64.9\%--98.5\% of LLM throughput. For example, on AGX Orin, robotic tracking misses are reduced by 96.1\% while retaining 93.4\% throughput; on OnePlus 13, camera deadline misses are reduced by 91.3\% while retaining 90.4\% throughput. Crucially, \system delivers all LLM outputs within their freshness windows. 
Overall, \system achieves our goal of preserving both timely sensor processing and useful LLM inference across distinct environments and workflows.  

\section{Background and Motivation}

\subsection{Edge Platforms with Unified Memory}
\begin{figure}[t]
    \centering
    \includegraphics[width=0.8\linewidth, trim=10mm 6mm 10mm 3mm]{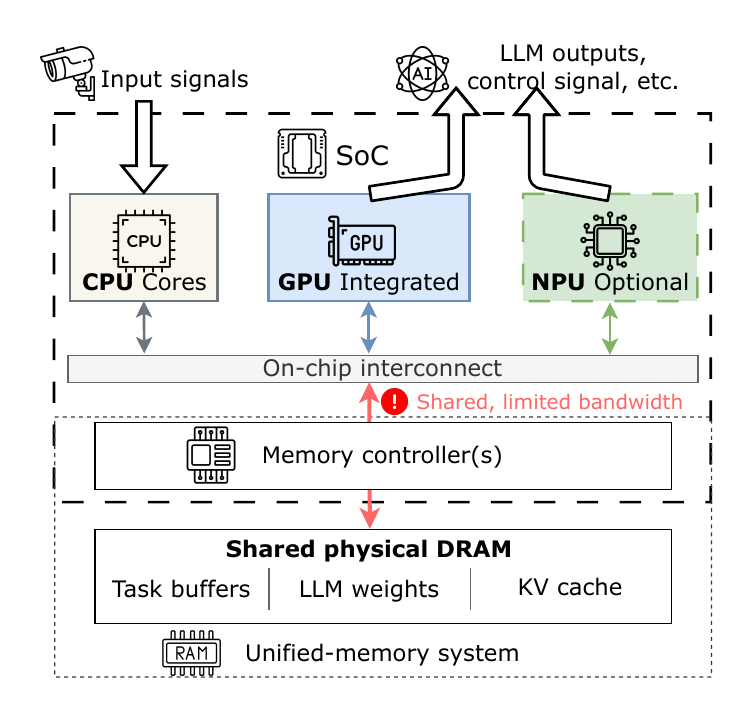}
    \caption{Edge platform with unified memory.
    CPU, GPU, and optional NPU traffic shares finite bandwidth to physical
    DRAM through the interconnect and the controller(s) of the unified-memory system. Task buffers, model weights, and the KV cache illustrate
    co-resident data. Caches and platform-specific channel organization
    are omitted.}
    \Description{CPU cores, an integrated GPU, and an optional NPU connect
    bidirectionally to an on-chip interconnect. A unified-memory system
    groups memory controllers and shared physical DRAM. A highlighted
    connection denotes their shared, finite memory bandwidth.}
    \label{fig:unified-memory-platform}
\end{figure}

Modern edge devices, from smartphones to NVIDIA Jetson boards, integrate CPU cores, GPUs, and, on many smartphones, a dedicated NPU into a single SoC sharing physical DRAM (Figure~\ref{fig:unified-memory-platform}). Typically, the CPU handles application logic and input processing, while accelerators execute neural-network computations~\cite{xu2025llmnpu,jetson}. 
Although the CPU and accelerators have separate execution resources, all of them will access DRAM through a common on-chip interconnect and memory controllers. This \emph{unified-memory system} reduces copies between processing units, allowing CPU-produced inputs to be consumed by an accelerator without a separate device-memory transfer~\cite{yi2025edgemoe}. It also means that the CPU task buffers, LLM model weights, intermediate activations, and the KV cache share the same physical memory. 

Efforts on model compression have successfully allowed the LLM and foreground state to fit in DRAM~\cite{ji2024awq,frantar2023gptq}, but capacity is not the only constraints. When LLM inference and CPU tasks run concurrently, their memory requests compete for memory bandwidth even though they use different address spaces and separate compute units. LLM traffic can therefore delay other tasks' memory accesses and inflate latency while CPU and accelerator compute resources remain available~\cite{xin2026sereno,wang2025agile3d,Esmaili2024}.


\subsection{Edge LLM Workflows and System Metrics}
A conventional interactive LLM request follows a prompt--generation--response sequence: the model processes an input and generates tokens for a user. Metrics including time to first token, token throughput, and request completion latency describe how quickly the response becomes available~\cite{agrawal2024sarathi}. In the edge applications we study here, LLM inference is additionally involved in a recurring observation--processing--inference--feedback loop. We refer to the CPU tasks that acquire and process sensor data as \emph{foreground tasks}, and the concurrently running LLM inference as the \emph{background task}. Foreground tasks 
provide observations to the background LLM, and may use its outputs to guide subsequent processing or actions. As new observations continue to arrive during inference, both must make progress concurrently on the same device.

This workflow introduces two distinct timing requirements. First, \textbf{input deadlines} require each foreground job to finish processing its observation within a budget measured from scheduled release, including queueing and execution. For example, a camera input task with a one-frame budget at 30\,Hz has about 33\,ms to complete. Missing this deadline can drop a frame, delay LLM inputs, or leave perception and control operating on stale state~\cite{zhang2019heteroedge,toba2024deadline}. Foreground processing must therefore continue on its own schedule while LLM inference runs. Second, \textbf{semantic freshness} requires an LLM to generate outputs timely, \emph{i.e.,} while its source observation remains useful. Its task-dependent validity window starts at the source observation and includes input preparation, waiting, inference, and delivery; an expired result may be discarded despite successful inference~\cite{badia2025age}. We target millisecond-scale input deadlines set by sensor cadence and seconds-scale LLM freshness windows set by how long the resulting observations remain relevant to the task. 

Correspondingly, we adopt the following metrics in the evaluation to fully cover the workflow requirements and characteristics. Foreground \emph{deadline miss rate} and \emph{response-time} measure whether observations are processed on time, reflecting the emphasis on deadline compliance in real-time task execution~\cite{toba2024deadline}. Following the use of information age to assess timely edge inference~\cite{badia2025age}, we consider \emph{result age at delivery} and additionally measure the \emph{fraction of outputs delivered within their freshness window}. LLM \emph{token throughput} and \emph{completion latency} quantify the inference service retained under concurrent execution. 
Finally, \emph{task-specific quality} captures whether timely execution produces useful application outcomes;

\subsection{Existing Solutions and Remaining Gaps}

Prior systems provide ways to control LLM inference, but their suspension points and deployment requirements differ from our settings. Sarathi-Serve divides prefill into chunks without providing comparable interruption points during decode~\cite{agrawal2024sarathi}. Finer control may rely on accelerator-specific preemption support or changes to the inference program or execution graph that must be maintained as models and runtimes evolve. Protecting short foreground deadlines calls for frequent, safe suspension points across both prefill and decode that preserve completed computation without depending on a particular accelerator.

SERENO is particularly relevant because it yields memory bandwidth to foreground mobile applications during LLM inference~\cite{xin2026sereno}. It creates yield points through speculative decoding, tying its control mechanism to a draft-and-verify workflow. Its foreground applications also run independently of the LLM. In the workflows we study, foreground tasks provide observations to the LLM and may use its outputs for subsequent actions. Holding inference can therefore protect input deadlines but also make the eventual feedback stale.

Finally, deciding \emph{when} to yield requires more than an aggregate bandwidth measurement, which does not reveal whether foreground execution is stalled. Detailed event tracing offers greater visibility but can interfere with the tasks being protected~\cite{zou2026profinfer, chu2025einfer}. The runtime needs actionable memory-pressure feedback at low cost, then must combine that feedback with foreground deadlines and LLM output freshness.

\subsection{Goals and Challenges}
\label{sec:challenges}


We seek a runtime that regulates background LLM execution so foreground tasks meet their input deadlines and LLM feedback arrives before its source observation becomes stale. Within those timing constraints, the runtime should preserve as much LLM progress as possible. Meeting these goals on a unified-memory edge device raises three challenges.

\vspace{0.05in}
\noindent\textbf{Challenge 1: Fine-grained and non-invasive LLM interruption.}
Foreground deadlines leave little time to react to rising memory pressure. As prior work on inference preemption illustrates~\cite{agrawal2024sarathi,xin2026sereno}, the choice of suspension point determines how quickly execution can yield and what overhead that control incurs. Pausing only after a long computation completes may relieve contention too late; accelerator work already submitted also continues after a hold is requested. More frequent suspension points shorten this delay but can increase scheduling and execution overhead. They must occur at safe boundaries so that model outputs remain unchanged and completed computation is preserved across a pause. The challenge is to provide such control during both prefill and decode, whose execution patterns differ, without extensive changes to the inference framework. 

\vspace{0.05in}
\noindent\textbf{Challenge 2: Accurate feedback on foreground memory pressure.}
High memory-bandwidth usage intuitively suggests strong contention and a higher foreground deadline-miss rate. However, bandwidth monitors report usage averaged over a sampling interval (e.g., 20\,ms), which can smooth out short bursts of memory demand. A burst may delay foreground execution even when the reported average stays below a warning threshold.
Our microbenchmarks expose the limitations of bandwidth-based warnings: for the camera workload, a 28\,GB/s threshold detects only 45.26\% of actual misses, while 93.84\% of its warnings are false positives. Lowering the threshold improves detectionm but most warnings still remain false (\cref{sec:eval-microbenchmarks}). The challenge is to identify signals that capture foreground memory demand and blocking with high accuracy to guide proper scheduling.

\begin{figure}[t]
    \centering
    \includegraphics[width=0.82\linewidth, trim=0mm 5mm 0mm 0mm]{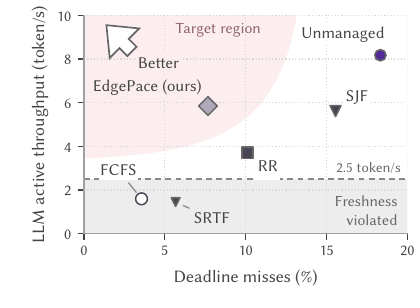}
    \caption{Four-camera scheduling on Jetson Orin Nano Super.
    Upper left favors fewer misses and higher throughput. Light red shading
    shows a schematic target; gray shading below 2.5\,token/s marks
    insufficient throughput for freshness. Unmanaged is basic \texttt{llama.cpp};
    RR is round-robin; SRTF is shortest remaining time first.}
    \Description{A scatter plot compares six scheduling policies. Deadline-miss
    rate increases to the right and throughput increases upward; an arrow in
    the upper-left corner points toward fewer misses and higher throughput.
    Unmanaged is purple, the other baselines are dark gray, and EdgePace is
    muted lavender. A light red upper-left region labeled Target region
    illustrates the desired combination of fewer misses and high throughput;
    its curved boundary is illustrative, not a fitted performance frontier.
    Unmanaged, FCFS, RR, SRTF,
    SJF, and EdgePace have miss rates of 18.33, 3.556, 10.111, 5.667,
    15.556, and 7.667 percent, respectively, and active throughputs of 8.1750,
    1.5976, 3.7088, 1.4483, 5.6149, and 5.8480 tokens per second.
    A dashed horizontal line marks 2.5 tokens per second; the gray region
    below it is labeled Freshness violated. FCFS and SRTF lie in this region
    and have separate leader labels; all their LLM outputs expire.}
    \label{fig:scheduler-comparison}
\end{figure}

\vspace{0.05in}
\noindent\textbf{Challenge 3: Scheduling for deadlines and LLM freshness.}
Classical scheduling rules alone do not account for foreground deadline urgency or LLM output freshness. In the four-camera experiment on Jetson Orin Nano Super (Figure~\ref{fig:scheduler-comparison}), first-come, first-served (FCFS) limits foreground misses to 3.56\% but leaves the LLM only 1.60\,token/s, below the 2.5\,token/s freshness requirement; all its outputs expire. Conversely, shortest job first (SJF) sustains 5.61\,token/s but leaves 15.56\% of foreground deadlines unmet. The root cause is that classic algorithms may prioritize fairness or service availability but lack visibility into application-side SLOs. Our goal is to reduce foreground misses as far as possible while retaining most LLM throughput to keep outputs fresh, moving toward the shaded upper-left region. This requires combining memory-pressure feedback with foreground deadline slack, LLM progress, and remaining freshness budgets. The scheduler must reserve time for outstanding work to drain and limit each pause so the LLM can still finish before its output expires.

\section{System Design}
\label{sec:design}

\system coordinates foreground latency-critical tasks and background LLM inference on unified-memory edge platforms. Its objective is to protect foreground deadlines while allowing the LLM to complete within the freshness lifetime of its input. This requires controlling both prefill and decode: delaying inference can relieve memory contention, but prolonged suspension can also render its output stale.

\begin{figure}[t]
  \centering
  \includegraphics[width=\linewidth, trim=0mm 5mm 0mm 7mm]{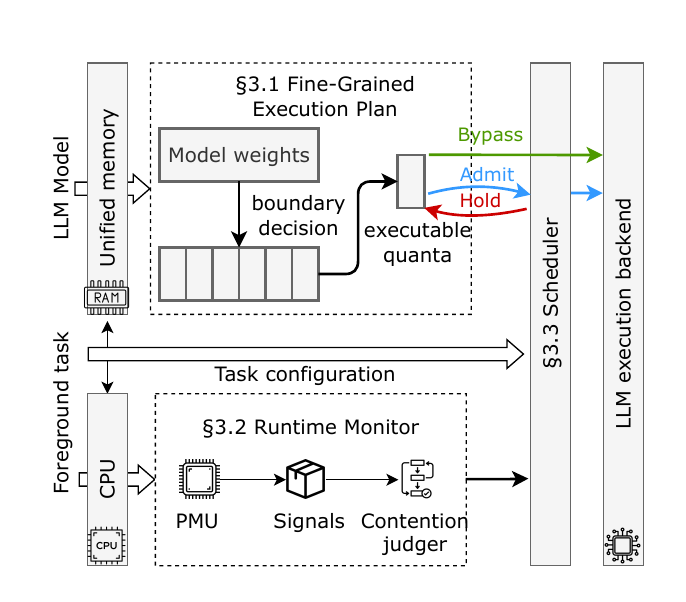}
  \caption{\system overview. The Fine Plan supplies legal execution
  boundaries and profiled runtimes. The monitor publishes PMU-derived demand
  and stall indicators. The scheduler combines them with task timing and
  LLM progress to control admission; the backend executes permitted quanta
  or continues bounded normal submission when no intervention is required.}
  \Description{The Fine Plan partitions LLM execution into legal quanta.
  The runtime monitor obtains foreground memory-demand and stall indicators
  from PMU counters. The scheduler receives these indicators and task state
  to select bypass, admit, or hold for LLM submissions. The backend executes
  permitted work and tracks completion. 
  }
  \label{fig:architecture}
\end{figure}

\vspace{0.05in}
\noindent \textbf{Design overview.} 
To tackle the challenges raised in \cref{sec:challenges}, 
\system mainly comprises three parts: \ding{182} an adaptive Fine Plan (\cref{sec:fine-plan}), \ding{183} a runtime monitor (\cref{sec:runtime_monitor}), and \ding{184} a risk- and freshness-aware scheduler (\cref{sec:scheduler}), as illustrated in Figure~\ref{fig:architecture}. The \textbf{Fine Plan} divides LLM execution at safe suspension boundaries. Each segment between successive boundaries is a \emph{quantum}; the plan records its expected execution time so the scheduler can account for work that may run before the next opportunity to pause. 
The \textbf{Monitor} uses Performance Monitoring Units (PMUs)~\cite{nvidia2026pmu,qualcommPerformanceTools} counters to estimate memory demand and stalls. The \textbf{Scheduler} combines these indicators with task timing and LLM progress to decide when submission should continue, become protected, or stop temporarily. At each safe boundary, the scheduler checks contention risk and freshness urgency to select \textsc{Bypass}, \textsc{Admit}, or \textsc{Hold}. \textsc{Bypass} preserves ordinary submission, \textsc{Admit} explicitly permits the next quantum, and \textsc{Hold} stops new submissions and periodically repeats both checks.

\subsection{Adaptive Fine-Grained Execution Plan}
\label{sec:fine-plan}
The Fine Plan determines where the LLM execution can pause to relieve shared-memory bandwidth contention. It divides the execution graph into scheduling quanta---segments that the backend can submit separately while preserving dependencies---and profiles their runtimes. The process comprises three steps.
\ding{182} \system identifies legal execution boundaries and profiles each quantum's runtime.
\ding{183} It selects safe sublayer boundaries for finer scheduling control, retaining runtime profiles specific to inference phase and execution progress.
\ding{184} It splits the graph into executable quanta and tracks their completion while preserving dependencies.


\vspace{0.05in}
\noindent \textbf{Discovering and profiling legal boundaries.} On first use of a model configuration, \system identifies the finest legal boundaries exposed by its backend adapter. Adjacent boundaries define scheduling quanta. Representative prompts provide each quantum's maximum observed submit-to-completion runtime, measured outside steady-state evaluation. These empirical estimates guide admission but are not worst-case execution-time bounds. Plans are cached by model and execution configuration, independently of priority and SLO settings. For Qwen2.5-1.5B on Jetson Orin Nano Super, initial profiling takes 9.204\,s, whereas cache reuse takes 20.787\,ms, motivating reuse across requests (\cref{sec:eval-microbenchmarks}). For Qwen-8B on Jetson AGX Orin, initial
profiling takes 12.06\,s, cache reuse just takes 54\,ms. Loading validates boundary order and positive, finite runtimes; all legal boundaries remain available for control. Adapting to a new LLM requires an adapter that identifies and profiles its legal execution boundaries.

\vspace{0.05in}
\noindent\textbf{Selecting suspension granularity.} Transformer execution offers several safe suspension boundaries: $(i)$ after a complete layer comprising attention, residual connections, normalization, and a feed-forward network (FFN); $(ii)$ after each token; or $(iii)$ within a layer, when a safely executable subgraph completes. We evaluate all three choices because the execution time between boundaries determines how frequently the scheduler can intervene. For Qwen2.5-1.5B, layer-level quanta measured across both prefill and decode average 44.14\,ms (35.27\,ms at p50 and 88.99\,ms at p95), while token-level quanta, which occur only during decode, average 21.26\,ms (7.88\,ms at p50 and 75.57\,ms at p95). Sublayer quanta measured across both phases average 2.26\,ms and reach 4.9\,ms at p95. These measurements motivate selecting safe sublayer boundaries: finer quanta provide more frequent opportunities to suspend execution and greater flexibility to fit LLM work into available foreground slack. Runtime profiles remain specific to prefill and decode, with separate profiles for the first and subsequent decode tokens.

\vspace{0.05in}
\noindent \textbf{Constructing executable quanta.} After device assignment and cross-backend copy handling, the backend adds graph splits at Fine Plan boundaries, preserving dependencies and intermediate tensors. Each quantum records its phase, identifier, and predicted runtime. Suspension stops new submissions while retaining completed computation; it does not interrupt executing kernels. A FIFO completion ring tracks outstanding backend events, distinguishing submitted work from completed service. Completing the final decode quantum advances token progress and selects the next runtime profile. For Qwen2.5-1.5B with 512 prefill and 128 decode tokens, partitioning increases total execution time from 7.39 to 7.70\,s, an increase of 4.2\% (Table~\ref{tab:micro-partition}). These measurements motivate evaluating partitioning cost alongside pause responsiveness.


\subsection{Runtime Monitoring of Memory Pressure}
\label{sec:runtime_monitor}

To address Challenge~2, the monitor uses four PMU counters on foreground CPU threads to capture both memory demand and execution already blocked on memory. We first explain their selection and combination into normalized indicators, then describe counter collection and publication. The scheduler combines these observations with task timing to assess deadline risk in \cref{sec:scheduler}.

\vspace{0.05in}
\noindent\textbf{Selecting and combining demand and stall signals.} Challenge~2 (\cref{sec:challenges}) identifies the limitations of interval-averaged bandwidth for detecting contention that threatens foreground deadlines. We therefore observe both foreground memory demand and the stalls already delaying execution.

For demand, we select last-level cache (LLC) read misses: a miss requires fetching data from beyond the LLC, making miss activity a proxy for memory-access demand. More requests can intensify bandwidth contention and delay foreground work. We measure misses both per second and per thousand retired instructions (MPKI). The time-based rate captures absolute request intensity, while the instruction-based metric captures miss density relative to completed work. Retired instructions therefore provide the normalization needed to distinguish memory-intensive execution from high activity alone.

For blocking, we select memory-related backend stall cycles, which reflect instruction execution impeded by memory waits. We normalize them by CPU cycles to measure the fraction of observed cycles spent stalled. Persistent stalls consume foreground timing slack and can lead to deadline misses. This signal complements demand: frequent accesses may be served without substantial blocking, whereas severe stalls can reduce the rate of further accesses. The four counters thus capture both demand and its observed effect on execution.

Over window $k$ of duration $\Delta t_k$, the monitor samples increments in CPU cycles $C_k$, retired instructions $I_k$, LLC read misses $M_k$, and memory-related backend stall cycles $B_k$. For a valid window, it computes cache misses per second $r_k$, MPKI $m_k$, and the memory-stall fraction $s_k$:
\begin{equation}
\label{eq:pmu-rates}
  r_k=\frac{M_k}{\Delta t_k},\qquad
  m_k=\frac{1000M_k}{I_k},\qquad
  s_k=\frac{B_k}{C_k}.
\end{equation}
Windows with nonpositive duration, zero instruction or cycle counts, or invalid/reset counters cannot supply these ratios and are marked invalid.

Let $\mathcal C(x)=\min\{1,\max\{0,x\}\}$ clip a value to $[0,1]$. Positive reference scales $r_0$, $m_0$, and $s_0$ normalize the three metrics into demand $D_k$ and stall $S_k$:
\begin{equation}
\label{eq:pmu-pressure}
\begin{aligned}
 D_k &= \mathcal C\!\left(\min\left\{\frac{r_k}{r_0},\frac{m_k}{m_0}\right\}\right),\\
 S_k &= \mathcal C\!\left(\frac{s_k}{s_0}\right).
\end{aligned}
\end{equation}
For cache-miss rate and MPKI, each reference scale is the larger of its historical baseline p99 and twice its p95; $s_0$ is the configured stall reference. Taking the minimum requires both absolute cache-miss intensity and misses per instruction to be high before demand is high, reducing false alarms from small instruction counts during low activity. Stall is retained independently: severe stalls can suppress completed memory accesses, so low demand can coexist with substantial memory-related blocking.
We smooth the two indicators independently, using $X_k$ to denote either $D_k$ or $S_k$:
\begin{equation}
\label{eq:pmu-smoothing}
 \bar X_k=\alpha X_k+(1-\alpha)\bar X_{k-1},
 \quad X\in\{D,S\},\quad 0<\alpha\leq1.
\end{equation}
The monitor publishes $(\bar D_k,\bar S_k)$, two dimensionless indicators in $[0,1]$ that summarize foreground memory demand and memory-related blocking. Larger values indicate stronger demand or stalls relative to their reference scales. These observations do not by themselves attribute stalls to the background LLM. The scheduler combines them with task timing in \cref{sec:scheduler}.

\vspace{0.05in}
\noindent\textbf{Counter collection and publication.} To reduce repeated acquisition, a separate producer samples counters at a configurable interval and publishes them to a fixed-size memory-mapped page. Scheduler checks reuse the latest valid snapshot between updates, keeping counter acquisition outside the submission path. A sequence counter ensures consistent reads; timestamps and validity/reset flags prevent stale or invalid samples from appearing as low pressure. On Orin Nano Super, median PMU reads consume 4.04\% and 0.65\% of the 2 and 20\,ms sampling intervals, respectively (Table~\ref{tab:micro-monitor-overhead}). These fractions quantify sampling cost, rather than application slowdown.

\subsection{Risk- and Freshness-Aware Scheduling}
\label{sec:scheduler}

The scheduler follows three steps: collect task state and compute timing budgets, check contention risk and freshness urgency at each legal boundary, and select an execution mode. The policy selects \textsc{Bypass}, \textsc{Admit}, or \textsc{Hold} for low, medium, or high risk, respectively, with freshness urgency overriding the choice to \textsc{Admit} (Algorithm~\ref{alg:scheduling}, lines~\ref{algline:freshness}--\ref{algline:hold-decision}). The backend enforces these decisions while preserving graph dependencies and completed computation and tracking progress (lines~\ref{algline:hold-stop}--\ref{algline:progress}). Only \textsc{Hold} activates periodic rechecking.

\vspace{0.05in}
\noindent\textbf{1. Collect task state and compute timing budgets.} Task timing determines how much delay foreground processing and LLM inference can tolerate. In Algorithm~\ref{alg:scheduling}, lines~\ref{algline:collect}--\ref{algline:budgets} read shared state and compute the available budgets. Foreground producers publish releases, deadlines, completion state, and historical response-time estimates. Request state supplies the LLM's source timestamp, freshness deadline, inference phase, and completed and remaining work. These task/request pages complement the PMU snapshots from \cref{sec:runtime_monitor}, keeping data collection outside the submission path.

For an active foreground job, the scheduler estimates its remaining work and computes two timing budgets:
\begin{equation}
\label{eq:scheduling-budgets}
\begin{aligned}
 \widehat W_{\mathrm{fg}} &= \max\{0,\widehat R_{\mathrm{fg}}-(t-r_{\mathrm{fg}})\},\\
 L_{\mathrm{fg}} &= (d_{\mathrm{fg}}-t)-\widehat W_{\mathrm{fg}},\\
 L_{\mathrm{fresh}} &= (d_{\mathrm{fresh}}-t)-\widehat S_{\mathrm{llm}}.
\end{aligned}
\end{equation}
Here, $t$ is the current time, and $r_{\mathrm{fg}}$ and $d_{\mathrm{fg}}$ are the foreground job's release time and deadline. The LLM output's freshness deadline is $d_{\mathrm{fresh}}$, determined from its source observation. All timestamps use a common monotonic clock. The historical foreground response-time estimate $\widehat R_{\mathrm{fg}}$ includes queueing and execution; $\widehat W_{\mathrm{fg}}$ estimates the time still needed to finish that job. The estimate $\widehat S_{\mathrm{llm}}$ covers remaining LLM service, including outstanding work, plus a reserve for parsing and delivery. Finally, $L_{\mathrm{fg}}$ is the foreground slack after reserving its remaining work, and $L_{\mathrm{fresh}}$ is the additional delay the LLM can tolerate before its output expires. A negative budget indicates insufficient remaining time. Both budgets are refreshed after every hold interval.

\vspace{0.05in}
\noindent\textbf{2. Check risk and freshness at execution boundaries.} Before the first quantum and at each subsequent legal boundary, the scheduler refreshes its inputs and produces two outputs: a contention-risk level $R$ and a freshness-urgency flag $F$ (Algorithm~\ref{alg:scheduling}, lines~\ref{algline:boundary}--\ref{algline:check}). Completion tracking updates the remaining-work estimates for subsequent checks (line~\ref{algline:progress}). Using the PMU indicators from \cref{sec:runtime_monitor}, the scheduler computes:
\begin{equation}
\label{eq:deadline-risk}
\begin{aligned}
 U(t) &= \mathcal C\!\left(1-\frac{L_{\mathrm{fg}}}{H}\right),\\
 \rho(t) &= w_D\bar D_k+w_S\bar S_k+w_U U(t),\\
 F(t) &= \mathbf{1}\{L_{\mathrm{fresh}}\leq 0\}.
\end{aligned}
\end{equation}
Here, $\mathcal C(\cdot)$ clips its input to $[0,1]$, and $H>0$ is the warning horizon in the same time units as $L_{\mathrm{fg}}$. The resulting $U(t)$ is normalized foreground deadline urgency. The index $k$ identifies the latest valid PMU sample; $\bar D_k$ and $\bar S_k$ are its smoothed memory-demand and stall indicators from \cref{sec:runtime_monitor}. The configured weights $w_D$, $w_S$, and $w_U$ control the contributions of demand, stalls, and urgency to the contention score $\rho(t)$. The indicator $\mathbf{1}\{\cdot\}$ returns one when its condition holds and zero otherwise, so the freshness-urgency flag $F(t)$ becomes one when no waiting budget remains, calling for immediate LLM progress.

The lower and upper risk thresholds $\theta_{\ell}<\theta_h$ map $\rho(t)$ to the discrete risk level $R$: low below $\theta_{\ell}$, medium in $[\theta_{\ell},\theta_h)$, and high at or above $\theta_h$. The score is a scheduling indicator, not a deadline-miss probability or rate. Every check evaluates both $R$ and $F$, including checks during \textsc{Bypass} and \textsc{Hold}.

The measured local checkpoint check averages 1.856\,µs and takes 7.135\,µs at p99 (\cref{sec:eval-microbenchmarks}), excluding PMU acquisition and pause/resume latency.


\begin{figure}[t]
    \centering
    \includegraphics[width=\linewidth]{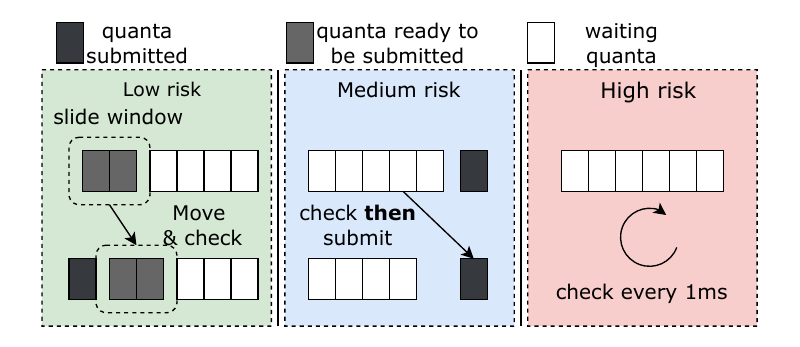}
    \caption{Risk-dependent LLM submission. Each block represents a
    \emph{quantum}: a segment of the LLM execution graph between legal
    pause boundaries that can be submitted separately, such as a layer
    or sublayer (\cref{sec:fine-plan}). 
    Low risk advances the submission
    window with boundary checks (\textsc{Bypass}); medium risk checks and
    submits one quantum (\textsc{Admit}); high risk defers submissions
    (\textsc{Hold}) and rechecks every 1\,ms (configurable). All checks evaluate freshness
    urgency, which overrides risk and selects \textsc{Admit}.}
    \Description{Three panels illustrate low-, medium-, and high-risk
    submission. Each block is a quantum, a separately submittable segment
    of the LLM execution graph between legal pause boundaries, such as a
    layer or sublayer. Dark blocks represent submitted quanta, medium-gray blocks
    are ready for submission, and light-gray blocks are waiting. At low
    risk, a sliding window advances with boundary checks. At medium risk,
    each quantum is checked before submission. At high risk, quanta remain
    pending while the scheduler checks periodically.}
    \label{fig:scheduling-policy}
\end{figure}

\vspace{0.05in}
\noindent\textbf{3. Select and enforce Bypass, Admit, or Hold.} Figure~\ref{fig:scheduling-policy} illustrates the three submission modes. The decision first considers freshness urgency (Algorithm~\ref{alg:scheduling}, lines~\ref{algline:freshness}--\ref{algline:hold-decision}). If $F$ is true, the scheduler selects \textsc{Admit} regardless of the risk level. Otherwise, low risk selects \textsc{Bypass}: after the boundary check, the backend continues ordinary submission without an explicit admission step or additional synchronization (line~\ref{algline:bypass-submit}). Medium risk selects \textsc{Admit}: the policy explicitly permits the next quantum and the backend submits it, with another check at the next boundary (line~\ref{algline:admit-submit}).

High risk selects \textsc{Hold}, stopping new submissions while allowing outstanding work to drain. The next quantum remains pending, and the scheduler repeats the state, budget, risk, and freshness checks every configurable interval $\delta$, set to 1\,ms by default (lines~\ref{algline:hold-stop}--\ref{algline:hold-recheck}). This polling loop runs only while held. A drop to low or medium risk selects \textsc{Bypass} or \textsc{Admit}, respectively; freshness urgency selects \textsc{Admit} even if risk remains high. Waiting therefore consumes the freshness budget and can itself cause a transition out of \textsc{Hold}.
Across different model measured \texttt{Admit} events, the next subgraph is submitted after 2.689\,ms at P99 (Table~\ref{tab:micro-transitions}). In our evaluation, \texttt{Admit} events mean account for only about 20\% of the total events, with the majority of events being \texttt{Bypass} events that incur no additional overhead; the brief pauses introduced by \texttt{Admit} events also alleviate contention pressure.


\begin{algorithm}[t]
\caption{Three-mode LLM scheduling}
\label{alg:scheduling}
\small
\begin{algorithmic}[1]
\Require Fine Plan $P$; shared state; hold interval $\delta$ (default 1\,ms)
\While{$P$ has an unsubmitted quantum}
    \State at a legal boundary, $q\gets$ next quantum\label{algline:boundary}
    \Repeat
        \State $S\gets$ read task, request, and PMU state\label{algline:collect}
        \State compute $L_{\mathrm{fg}},L_{\mathrm{fresh}}$ from $S$ and current time\label{algline:budgets}
        \State $(R,F)\gets$ risk and freshness check\label{algline:check}
        \If{$F$}\label{algline:freshness}
            \State $a\gets\textsc{Admit}$
        \ElsIf{$R=\text{low}$}
            \State $a\gets\textsc{Bypass}$
        \ElsIf{$R=\text{medium}$}
            \State $a\gets\textsc{Admit}$
        \Else
            \State $a\gets\textsc{Hold}$\label{algline:hold-decision}
        \EndIf
        \If{$a=\textsc{Hold}$}
            \State stop new submissions; let outstanding work drain\label{algline:hold-stop}
            \State wait $\delta$; keep $q$ pending\label{algline:hold-wait}
        \EndIf
    \Until{$a\neq\textsc{Hold}$}\label{algline:hold-recheck}
    \If{$a=\textsc{Bypass}$}
        \State continue ordinary submission of $q$\label{algline:bypass-submit}
    \Else
        \State explicitly admit and submit $q$\label{algline:admit-submit}
    \EndIf
    \State track submitted and completed work\label{algline:progress}
\EndWhile
\State wait for remaining work to complete
\end{algorithmic}
\end{algorithm}

\section{Evaluation}
\label{sec:evaluation}

\noindent\textbf{Implementation.}The external control module contains 1,702 lines of C++(LoC), including the ABI wrapper(242 LoC) and the policy engine(1,460 LoC). This code is fully external to the llama.cpp inference core and is loaded through a dynamic C ABI.  Only the backend scheduler dispatch path of llama.cpp is modified, adding approximately 290 LoC to classify execution phases, maintain quantum metadata, and emit submission/completion events. Model loading, tensor operators, graph construction, KV-cache management, and numerical kernels remain unchanged. In addition, risk phase identification requires calibration. After changing the platform and foreground load, a foreground task run without LLM and a total closed-loop task run without management are required to jointly calibrate the risk range.

Our evaluation combines microbenchmarks of monitoring accuracy and cost, decision overhead, pause/resume responsiveness, and graph partitioning (\cref{sec:eval-microbenchmarks}) with two end-to-end case studies: multi-camera semantic feedback (\cref{sec:eval-case1}) and MuJoCo-based robotic target tracking~\cite{todorov2012mujoco} (\cref{sec:eval-case2}). We measure foreground deadline misses, latency percentiles, LLM throughput, and task quality to assess timely foreground execution alongside LLM progress.

\subsection{Experimental Setup}
\label{sec:eval-setup}

\noindent\textbf{Testbed.}
Table~\ref{tab:eval-testbed} lists our four unified-memory testbeds and their memory configurations. The OnePlus 13 Android phone uses a Snapdragon 8 Elite SoC, and the other three Jetson platforms are equipped with ARM CPUs and integrated NVIDIA GPUs, each with different specs. The OnePlus phone runs Case Study~1; the three Jetsons run both cases.

\begin{table}[t]
    \centering
    \footnotesize
    \setlength{\tabcolsep}{3pt}
    \renewcommand{\arraystretch}{1.15}
    \caption{Evaluation testbeds and memory configurations.}
    \label{tab:eval-testbed}
    \begin{tabular*}{0.94\linewidth}{@{\extracolsep{\fill}}lrr@{}}
        \toprule
        {\sffamily\bfseries Testbed} &
        {\sffamily\bfseries Bandwidth (GB/s)} &
        {\sffamily\bfseries Capacity (GB)} \\
        \midrule
        Jetson Orin Nano Super~\cite{Jetson-Nano-Orion} & 102 & 8 \\
        Jetson Thor~\cite{nvidia2026jetsonThor} & 273 & 128 \\
        Jetson AGX Orin~\cite{jetson} & 204.8 & 64 \\
        OnePlus 13~\cite{oneplus13specs,qualcomm2024snapdragon8elite} & 84.8 & 24 \\
        \bottomrule
    \end{tabular*}
\end{table}

\noindent\textbf{Model selection.}
We match models to platform memory and inference capacity while reserving resources for foreground tasks. Orin Nano Super runs Qwen2.5-1.5B~\cite{qwen2024qwen25}; AGX Orin and Thor run Qwen3-8B~\cite{yang2025qwen3} and Qwen3-VL-8B-Instruct~\cite{bai2025qwen3vl}; OnePlus 13 runs Qwen2.5-1.5B and Qwen3-VL-2B-Q8. In both cases, text LLM and VLM requests arrive every 1.6\,s and 5\,s, respectively, with 246-token input contexts and at most 16 output tokens. These short outputs make vision processing a substantial part of VLM request time. Runtime comparisons use identical platform, model, and foreground settings; workloads differ across platforms to match their capacity.

\noindent\textbf{Inference baselines.}
\emph{Unmanaged} is basic \texttt{llama.cpp}~\cite{llamacpp} without
\system scheduling. Jetson baselines additionally include the
model compiler MLC-LLM~\cite{mlcllm} and NVIDIA's GPU inference
optimizer TensorRT~\cite{nvidiaTensorRT}; phone baselines include
the lightweight Mobile Neural Network (MNN)~\cite{jiang2020mnn}
and Snapdragon's Qualcomm AI Runtime (QAIRT) SDK~\cite{qualcommQairt}(all experiments on phone use the NPU).
In our tested configurations, MLC-LLM had issues with VLMs,
MNN with Jetson, and MLC-LLM failed using NPU on OnePlus 13.
TensorRT also ran out of memory (OOM) on Orin Nano Super.
We report only successfully completed combinations. The figure legends group MLC-LLM, TensorRT, MNN, and QAIRT under ``Alternative baseline''.


\noindent\textbf{Case-study workloads.}
Case Study~1 uses camera snapshots and asynchronous LLM feedback to select cameras and processing modes, measuring the quality of state consumed at foreground deadlines. Case Study~2 measures observation freshness in a MuJoCo Panda tracking task; concurrent model outputs do not alter tracking actions. The case studies specify workload settings, timing windows, and quality metrics; memory-pressure observations characterize contention.

\subsection{Microbenchmarks}
\label{sec:eval-microbenchmarks}

Unless stated otherwise, microbenchmarks use Orin Nano Super, our least powerful platform, to expose overheads and bottlenecks under tight resource constraints. Monitoring accuracy uses separate online-instrumented reruns
of both case-study workloads (\cref{sec:eval-case1,sec:eval-case2}).

\begin{figure}[t]
    \centering
    \includegraphics[width=\columnwidth, trim=0mm 4mm 0mm 0mm]{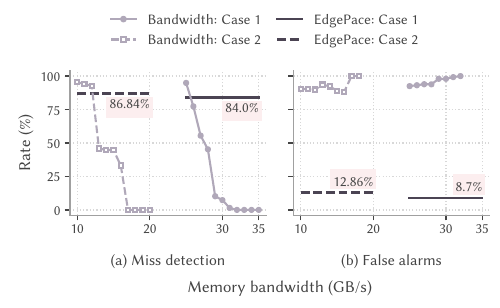}
    \caption{Monitoring accuracy on Orin Nano Super: (a) miss detection
    (higher is better) and (b) false alarms among warnings (lower is better).
    The x-axis shows memory bandwidth (GB/s), used as the baseline's
    warning threshold.
    Each panel compares Case~1 (multi-camera, solid) and Case~2
    (robotic tracking, dashed) over their measured threshold ranges.
    Dark horizontal references highlight \system; light curves show the
    bandwidth baseline. Unreported values are omitted.}
    \Description{The left panel compares miss detection for both cases;
    the right compares false-alarm fractions. Case 1 uses thresholds
    from 25 to 35 GB/s and Case 2 from 10 to 20 GB/s.
    Baseline detection decreases sharply as the threshold rises,
    while reported false-alarm fractions remain above 88 percent.
    EdgePace detects 84.0 and 86.84 percent of misses with false-alarm
    fractions of 8.7 and 12.86 percent in Cases 1 and 2, respectively.}
    \label{fig:monitor-thresholds}
\end{figure}

\noindent\textbf{Monitoring accuracy.}
Figure~\ref{fig:monitor-thresholds} compares \system with a monitor that
flags deadline misses when bandwidth exceeds a threshold.
Miss detection measures correctly flagged misses over all actual misses;
the false-alarm fraction measures incorrect warnings over all warnings.
\system detects 84.0\% and 86.84\% of misses in Cases~1 and~2, respectively,
with false-alarm fractions of 8.7\% and 12.86\%.
Its false-alarm fractions are 83.80 and 75.51 percentage points below
the baseline's lowest reported values, retaining high miss detection
with substantially fewer false warnings.

\vspace{0.05in}
\noindent\textbf{Monitoring and decision overhead.}
On Orin Nano Super, median PMU reads take 80.9 and 130.8\,$\mu$s
at 2 and 20\,ms sampling intervals, respectively
(Table~\ref{tab:micro-monitor-overhead}). These consume 4.04\% and
0.65\% of their intervals; the fractions measure neither application
slowdown nor total CPU utilization. Excluding PMU acquisition, the
local decision check averages 1.856\,$\mu$s, with P50/P95/P99 of
1.184/3.199/7.135\,$\mu$s: only 0.0056\% of a 33.333\,ms
foreground period and substantially less than a PMU read.

\begin{table}[t]
    \centering
    \footnotesize
    \setlength{\tabcolsep}{3pt}
    \renewcommand{\arraystretch}{1.15}
    \caption{PMU sampling cost on Orin Nano Super.
    Read/interval is the median read latency as a percentage of
    the sampling interval.}
    \label{tab:micro-monitor-overhead}
    \begin{tabular*}{0.94\linewidth}{@{\extracolsep{\fill}}rrrrr@{}}
        \toprule
        \multirow{2}{*}{\sffamily\bfseries Interval (ms)} &
        \multicolumn{3}{c}{\sffamily\bfseries Read latency ($\mu$s)} &
        \multirow{2}{*}{\sffamily\bfseries Read/interval (\%)} \\
        \cmidrule(lr){2-4}
        & {\sffamily\bfseries P50} & {\sffamily\bfseries P95} &
        {\sffamily\bfseries P99} & \\
        \midrule
         2 &  80.9 & 148.7 & 232.1 & 4.04 \\
         5 & 116.8 & 178.3 & 411.1 & 2.34 \\
        10 & 127.9 & 211.7 & 646.2 & 1.28 \\
        20 & 130.8 & 187.4 & 492.1 & 0.65 \\
        \bottomrule
    \end{tabular*}
\end{table}

\noindent\textbf{Hold and Admit responsiveness.}
Table~\ref{tab:micro-transitions} reports P99 latencies for
Qwen2.5-1.5B on Orin Nano Super and both 8B models on AGX Orin;
Appendix~\ref{sec:appendix-transitions} gives all P50/P95/P99 values
and sample counts. In all three configurations, \texttt{Hold}
stops submissions at a legal boundary with a recorded delay of 0\,ms,
but outstanding GPU work still drains: this is not kernel preemption.
Overall P99 drain latency is 3.202--4.176\,ms, with longer tails
in Nano decode (5.429\,ms) and AGX VLM vision processing
(4.724\,ms versus 2.756\,ms in decode).

Overall P99 resumption latency is 1.221--2.689\,ms from
\texttt{Admit} and 1.713--3.037\,ms from the end of \texttt{Hold}
to the next subgraph submission, excluding subsequent GPU queueing.
Effective pause/resume latency thus includes drain and resubmission
costs beyond the local decision check.

\begin{table}[t]
    \centering
    \footnotesize
    \setlength{\tabcolsep}{3pt}
    \renewcommand{\arraystretch}{1.15}
    \caption{P99 pause and resume delays (ms).
    $H$: \texttt{Hold} decision to drain completion;
    $A$: \texttt{Admit} decision to next submission;
    $R$: end of \texttt{Hold} to next submission.}
    \label{tab:micro-transitions}
    \begin{tabular*}{0.94\linewidth}{@{\extracolsep{\fill}}lrrr@{}}
        \toprule
        {\sffamily\bfseries Phase} &
        {\sffamily\bfseries $H$: drain} &
        {\sffamily\bfseries $A$: submit} &
        {\sffamily\bfseries $R$: submit} \\
        \midrule
        \multicolumn{4}{@{}l}{\sffamily\bfseries Orin Nano Super: Qwen2.5-1.5B} \\
        All     & 4.176 & 2.689 & 3.037 \\
        Prefill & 3.989 & 1.865 & 2.229 \\
        Decode  & 5.429 & 3.040 & 2.835 \\
        \midrule
        \multicolumn{4}{@{}l}{\sffamily\bfseries AGX Orin: Qwen3-8B} \\
        All     & 3.202 & 1.221 & 1.713 \\
        Prefill & 3.348 & 1.515 & 1.961 \\
        Decode  & 2.906 & 0.847 & 1.329 \\
        \midrule
        \multicolumn{4}{@{}l}{\sffamily\bfseries AGX Orin: Qwen3-VL-8B} \\
        All     & 3.719 & 1.546 & 1.806 \\
        Prefill & 3.371 & 1.774 & 2.176 \\
        Decode  & 2.756 & 0.694 & 1.125 \\
        Vision  & 4.724 & 0.617 & 0.975 \\
        \bottomrule
    \end{tabular*}
\end{table}

\noindent\textbf{Graph partitioning overhead.}
With 512 prefill tokens, 128 decode tokens, and a 512-token visual
budget for VLMs, partitioning increases total execution time by
0.25--4.50\% across platforms, including 4.19\% on Orin Nano Super
(Table~\ref{tab:micro-partition}). VLM totals include vision encoding;
Appendix~\ref{sec:appendix-partition} provides the phase breakdown.
Separately, initial Nano profiling for Qwen2.5-1.5B takes 9.204\,s;
reusing a cached plan takes 20.787\,ms and avoids repeated profiling.


\begin{table}[t]
    \centering
    \footnotesize
    \setlength{\tabcolsep}{2pt}
    \renewcommand{\arraystretch}{1.15}
    \caption{Total execution time with and without graph partitioning
    for 512 prefill tokens and 128 decode tokens; VLMs use a
    512-token visual budget, with vision encoding included in the total.
    Increases are calculated before rounding.}
    \label{tab:micro-partition}
    \begin{tabular*}{0.96\linewidth}{@{\extracolsep{\fill}}lrrr@{}}
        \toprule
        {\sffamily\bfseries Model} &
        {\sffamily\bfseries Unpartitioned (s)} &
        {\sffamily\bfseries Partitioned (s)} &
        {\sffamily\bfseries Increase (\%)} \\
        \midrule
        \multicolumn{4}{@{}l}{\sffamily\bfseries Orin Nano Super} \\
        Qwen2.5-1.5B & 7.39 & 7.70 & 4.19 \\
        \midrule
        \multicolumn{4}{@{}l}{\sffamily\bfseries AGX Orin} \\
        Qwen3-8B & 5.744 & 5.920 & 3.07 \\
        Qwen3-VL-8B & 7.553 & 7.572 & 0.25 \\
        \midrule
        \multicolumn{4}{@{}l}{\sffamily\bfseries Jetson Thor} \\
        Qwen3-8B & 3.824 & 3.842 & 0.46 \\
        Qwen3-VL-8B & 4.821 & 4.849 & 0.57 \\
        \midrule
        \multicolumn{4}{@{}l}{\sffamily\bfseries OnePlus 13} \\
        Qwen2.5-1.5B & 8.069 & 8.432 & 4.50 \\
        Qwen3-VL-2B-Q8 & 26.027 & 26.673 & 2.48 \\
        \bottomrule
    \end{tabular*}
\end{table}

\subsection{Case Study 1: Multi-Camera Semantic Feedback}
\label{sec:eval-case1}
Camera and region selection depend on foreground observations~\cite{liu2022multi} that may become stale under delayed processing~\cite{li2020towards}. We evaluate how deadline protection affects LLM throughput and selection quality.

\begin{figure*}[t]
    \centering
    \includegraphics[width=\textwidth, trim=0mm 2mm 0mm 0mm]{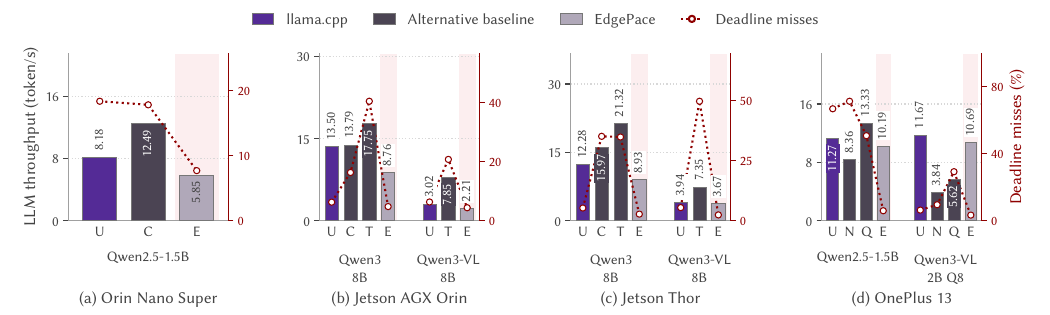}
    \caption{Multi-camera foreground deadline misses and LLM throughput on four platforms.
    Bars show token throughput (left axis); dark-red dotted lines show
    deadline-miss rates (right axis, lower is better).
    Lines connect runtimes within each model group.
    U = Unmanaged (basic \texttt{llama.cpp}), C = MLC-LLM, T = TensorRT,
    N = MNN, Q = QAIRT, and E = \system. Light red bands highlight \system.
    Axis scales
    and foreground loads vary by platform.}
    \Description{Four platform panels combine throughput bars with red deadline-miss
    lines for each model. EdgePace columns have light red backgrounds.
    EdgePace has the lowest miss rate in all seven configurations,
    with the accompanying token throughput shown separately.}
    \label{fig:case1-performance}
\end{figure*}

\begin{figure*}[t]
    \centering
    \includegraphics[width=\textwidth, trim=0mm 2mm 0mm 0mm]{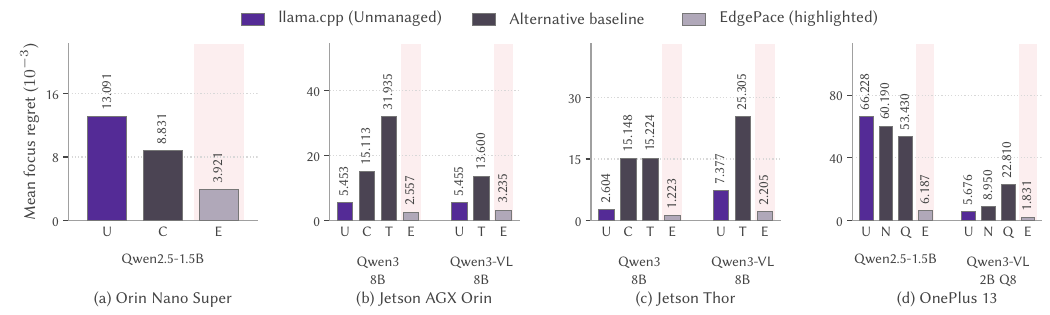}
    \caption{Task quality for the completed platform--model experiments in
    Figure~\ref{fig:case1-performance}, using the same method abbreviations.
    Mean focus regret averages regional selection loss over all epochs,
    including zero-loss epochs; lower is better. Values are in units of
    $10^{-3}$, with separate vertical scales per platform.
    Panel (a) uses the four-camera Nano measurements.
    Light red bands highlight \system. Qwen3-VL-8B denotes the Instruct model.}
    \Description{Four platform panels show mean focus regret for each recorded model
    and runtime. The Nano panel uses four-camera data: 13.091, 8.831,
    and 3.921 in units of one thousandth for llama.cpp, MLC-LLM, and EdgePace.
    EdgePace has the lowest regret in all seven configurations.}
    \label{fig:case1-quality}
\end{figure*}

\vspace{0.05in}
\noindent\textbf{Workload and baselines.}
Parallel camera pipelines replay an OpenEQA~\cite{majumdar2024openeqa} video subset at 30 Hz, computing activity and motion statistics from aligned frames each epoch under a one-frame deadline of 33.33 ms. They publish semantic snapshots to an asynchronous LLM; valid, fresh outputs select cameras and processing modes, otherwise defaults apply.

The four-camera microbenchmark load is heavy for Nano but light for AGX Orin, motivating workloads matched to platform decoding, memory, and compute capacity. Nano and OnePlus 13 use four 1080p streams, AGX Orin uses seven 1080p streams, and Thor uses four camera feeds with 30 decode streams. 


Figures~\ref{fig:case1-performance} and~\ref{fig:case1-quality} compare the completed runs of \system and the baselines in \cref{sec:eval-setup}, excluding the phone stress test with continuous co-execution. Comparisons are within platform--model pairs because foreground loads differ across platforms.


\vspace{0.05in}
\noindent\textbf{Task-quality metric.}
At each deadline, a separate consumer selects the most active of four regions using the latest completed foreground state. Stale state causes a \emph{focus mismatch} only when this selection differs from the current-frame reference. We report mismatch frequency over all epochs and mean focus regret:
\begin{equation}
\bar{r} = \frac{1}{N}\sum_{i=1}^{N}
\left(\max_z q_{i,z}-q_{i,\hat{z}_i}\right),
\label{eq:case1-focus-regret}
\end{equation}
where $N$ is the epoch count, $q_{i,z}$ is regional activity occupancy under the \texttt{strong \&\& high\_motion} rule, and $\hat{z}_i$ is the region selected at the deadline. The current frame supplies the scoring reference after completion. These metrics assess timely state consumption, not LLM prediction accuracy; stale state incurs no regret if it preserves the best region.

\vspace{0.05in}
\noindent\textbf{Deadline misses and token throughput.}
Across seven platform--model pairs, \system has the lowest measured miss rate, reducing misses by 24.1--91.3\% relative to Unmanaged while retaining 64.9--93.2\% of its throughput (Figure~\ref{fig:case1-performance}). Four-camera Nano misses fall from 18.33\% to 7.667\% (58.2\% fewer), with throughput decreasing from 8.175 to 5.848\,token/s (71.5\% retained). This demanding load leaves limited inference headroom. MLC-LLM reaches 12.49\,token/s at 17.78\% misses; \system reduces its miss rate by 56.9\%.

For AGX Orin's text and vision-language models, miss reductions are 24.1\% and 31.1\%, retaining 64.9\% and 73.2\% of unmanaged throughput. Thor achieves 48.9\% and 56.6\% reductions with 72.7\% and 93.2\% retention. Faster runtimes incur more misses: TensorRT reports 20.72--49.83\%, and the three MLC-LLM runs report 16.28--35.17\%. Against these baselines, \system reduces misses by 79.1--95.3\% and 56.9--92.4\%, retaining 28.2--49.9\% and 46.8--63.5\% of their respective throughput.

OnePlus 13 with Qwen2.5-1.5B achieves the largest miss reduction: 66.78\% to 5.83\% (91.3\%), retaining 90.4\% of unmanaged throughput at 10.188\,token/s. MNN and QAIRT incur 71.28\% and 50.70\% misses at 8.365 and 13.327\,token/s. With Qwen3-VL-2B-Q8, misses fall from 6.22\% to 3.22\%, retaining 91.6\% of unmanaged throughput at 10.693\,token/s. This improves both metrics over MNN (9.44\%, 3.838\,token/s) and QAIRT (29.17\%, 5.620\,token/s). For all methods, token throughput is computed as generated tokens divided by request arrival-to-completion time, including scheduling delays, and comparisons are made within configurations.

Appendix~\ref{sec:appendix-case1} combines this four-camera Nano result with lighter-load runs: the three-camera(medium-load in Appendix B) run retains 75.2\% of unmanaged throughput, versus 71.5\% with four cameras. The appendix's heavy-load point matches Figure~\ref{fig:case1-performance}.

\begin{figure*}[t]
    \centering
    \includegraphics[width=\textwidth, trim=0mm 2mm 0mm 0mm]{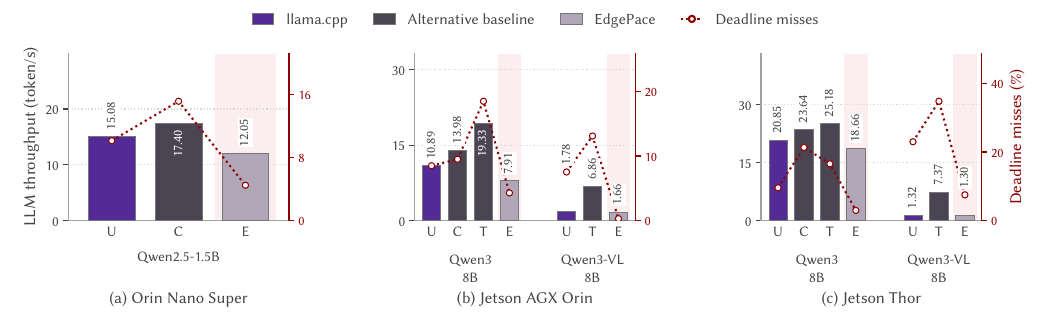}
    \caption{Robotic target-tracking deadline misses and LLM throughput on three
    Jetsons. Bars show token throughput (left axis); dark-red dotted lines
    show sensor deadline-miss rates (right axis, lower is better).
    Lines connect runtimes within
    each model group. U = Unmanaged (basic \texttt{llama.cpp}), C = MLC-LLM,
    T = TensorRT, and E = \system. Light red bands highlight \system.
    Each group contains its completed runtime experiments;
    axis scales differ by platform.}
    \Description{Three platform panels combine token-throughput bars and red sensor
    deadline-miss lines. EdgePace has lower miss rates than each measured
    runtime across all five configurations. Its columns have light red
    backgrounds, and the bars report the accompanying throughput.}
    \label{fig:case2-performance}
\end{figure*}

\begin{figure*}[t]
    \centering
    \includegraphics[width=\textwidth, trim=0mm 2mm 0mm 0mm]{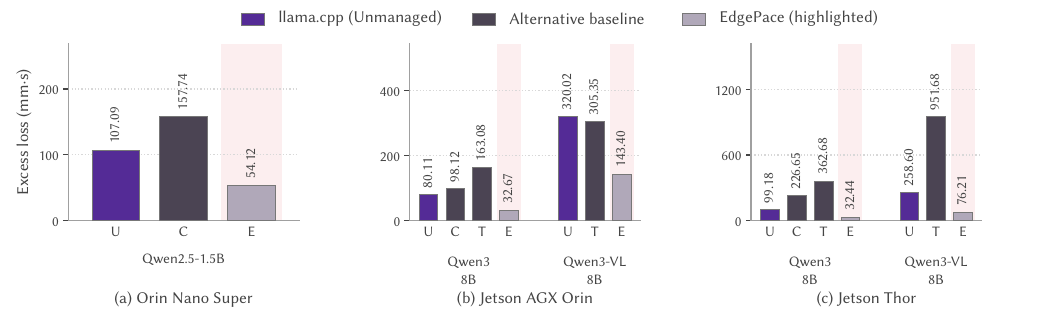}
    \caption{Observation freshness for the completed platform--model experiments
    in Figure~\ref{fig:case2-performance}, using the same method abbreviations.
    Bars show excess integrated staleness loss $\Delta S=S-S_0$ relative to
    the matching No-LLM run, in $\mathrm{mm}\cdot\mathrm{s}$; lower is better.
    Light red bands highlight \system. Vertical scales differ across platforms.}
    \Description{Three panels show excess integrated staleness loss for every recorded
    runtime and model, using a matching sensing-only reference.
    EdgePace has lower excess loss than each measured baseline
    in all five configurations.}
    \label{fig:case2-quality}
\end{figure*}

\vspace{0.05in}
\noindent\textbf{Effect on task quality.}
\system has the lowest mean focus regret in all seven configurations (Figure~\ref{fig:case1-quality}). Relative to Unmanaged, reductions are 70.0\% on four-camera Nano (0.013091 to 0.003921), 53.1\%/40.7\% on AGX Orin, 53.0\%/70.1\% on Thor, and 90.7\%/67.7\% on OnePlus 13 (text/VLM, respectively). Thor's VLM regret falls from 0.007377 to 0.002205. Against MLC-LLM's text-model runs, reductions are 55.6\%, 83.1\%, and 91.9\% on Nano, AGX Orin, and Thor, respectively; against TensorRT they are 76.2--92.0\%. Phone regret falls by 79.5--92.0\% relative to MNN and QAIRT. Thus, fewer misses accompany more reliable task-state consumption.

In the separate lightweight three-camera(light-load in Appendix B) Nano run, \system reduces stale consumptions from 186 to 106 and focus mismatches from 91 to 50. Mismatch frequency falls from 2.53\% to 1.39\%, while conditional mean regret among mismatches changes less (0.1007 to 0.0897), suggesting that fewer incorrect selections drive the benefit.

Additional Nano experiments on foreground-load and quantization sensitivity appear
in Appendix~\ref{sec:appendix-case1} (Figure~\ref{fig:case1-sensitivity}).

\subsection{Case Study 2: Robotic Target Tracking}
\label{sec:eval-case2}
This case tests whether protecting sensor processing preserves the freshness of robot-control observations while sustaining concurrent inference.

\vspace{0.05in}
\noindent\textbf{Workload and baselines.}
A Panda arm in MuJoCo~\cite{todorov2012mujoco} tracks a target moving along a horizontal circle of radius 4\,cm. Physics advances every 2\,ms; RGB-D sensing and control run every 50\,ms with a 40\,ms sensing-to-command deadline, leaving a 10 ms timing margin before the next control update. Simulator ground-truth target and end-effector states reach the controller only after RGB-D processing completes; until then, it reuses prior state. Concurrent model outputs alter neither the trajectory nor tracking actions, so quality measures consumed-state freshness.

We scale RGB-D resolution with rendering, memory, and compute capacity while keeping the workflow, control period, and deadline fixed: Nano runs Qwen2.5-1.5B at $400\times400$; AGX Orin and Thor run both 8B models at $540\times540$ and $656\times656$, respectively. Each platform--model pair compares \system and the inference baselines under identical foreground settings, with a matching No-LLM run as the sensing-only reference. Comparisons remain within pairs because rendering loads and reported timing windows differ across platforms.


\vspace{0.05in}
\noindent\textbf{Task-quality metric.}
At control time $t_i$, the consumed observation has source timestamp $s_i$; let $\ell_i$ be the target-position discrepancy between these times, in millimeters. Over the steady-state window,
\begin{equation}
S = \Delta t\sum_i \ell_i, \qquad
\Delta S = S-S_0, \qquad \Delta t=0.05\,\mathrm{s},
\label{eq:case2-staleness}
\end{equation}
where $S_0$ is the matching No-LLM result. Both total loss $S$ and excess loss $\Delta S$ use $\mathrm{mm}\cdot\mathrm{s}$ (lower is better). Subtracting $S_0$ isolates observation lag added by co-execution from sensing-only loss.

\vspace{0.05in}
\noindent\textbf{Deadline misses and token throughput.}
\system has the lowest measured miss rate across all five pairs, reducing misses by 49.6--96.1\% relative to Unmanaged while retaining 72.7--98.5\% of its throughput (Figure~\ref{fig:case2-performance}). Nano misses fall from 10.12\% to 4.47\% (55.8\% fewer), retaining 79.9\% of throughput (15.08 to 12.05\,token/s). MLC-LLM instead incurs 15.13\% misses at 17.40\,token/s.

For AGX Orin's Qwen3-8B and Qwen3-VL-8B, misses fall from 8.50\% to 4.28\% and from 7.53\% to 0.29\%, retaining 72.7\% and 93.4\% of throughput, respectively. Thor's corresponding miss rates fall from 9.56\% to 2.94\% and from 23.00\% to 7.50\%, with 89.5\% and 98.5\% retention. The remaining misses, particularly Thor's VLM result, show that scheduling mitigates rather than eliminates deadline violations.

TensorRT delivers higher throughput on AGX Orin and Thor but incurs 13.11--34.83\% misses. \system reduces these misses by 76.9--97.8\% at 17.7--74.1\% of its throughput. Thor's VLM has the largest throughput cost (1.301 versus 7.37\,token/s), with misses falling from 34.83\% to 7.50\%. MLC-LLM incurs 9.50\% misses at 13.978\,token/s for AGX Orin's text model and 21.33\% at 23.64\,token/s for Thor's. Across the three MLC-LLM comparisons, \system reduces misses by 54.9--86.2\% while retaining 56.6--78.9\% of throughput.

\vspace{0.05in}
\noindent\textbf{Effect on task quality.}
\system yields the lowest excess staleness loss in every configuration
(Figure~\ref{fig:case2-quality}). Relative to Unmanaged, Nano's $\Delta S$
falls from 107.09 to 54.12\,$\mathrm{mm}\cdot\mathrm{s}$ (49.5\%).
AGX Orin's text/VLM losses fall from 80.11 to 32.67 (59.2\%) and
320.02 to 143.40\,$\mathrm{mm}\cdot\mathrm{s}$ (55.2\%);
Thor's fall from 99.18 to 32.44 (67.3\%) and 258.60 to
76.21\,$\mathrm{mm}\cdot\mathrm{s}$ (70.5\%). These gains accompany
fewer misses, indicating fresher consumed observations. No-LLM runs
already incur roughly 809\,$\mathrm{mm}\cdot\mathrm{s}$ despite zero
sensor misses; hence total-loss reductions are 5.3--17.1\%, compared
with 49.5--70.5\% for excess loss.

MLC-LLM incurs 157.74, 98.12, and 226.65\,$\mathrm{mm}\cdot\mathrm{s}$
on Nano, AGX Orin, and Thor's text-model runs; \system reduces these
by 65.7\%, 66.7\%, and 85.7\%, respectively. TensorRT incurs
163.08--951.68\,$\mathrm{mm}\cdot\mathrm{s}$ across the four 8B
configurations, reduced by 53.0--92.0\% with \system. These runtimes'
higher throughput thus accompanies less timely state consumption.

Appendix~\ref{sec:appendix-risk} reports scheduler decision distributions
and cumulative \textsc{Hold} waiting times for both case studies.


\section{Related Work}
\label{sec:realated work}
\textbf{Model scheduling on edge devices.}
Prior systems improve on-device DNN execution through hardware-aware scheduling and resource-aware model adaptation.
$\mu$Layer~\cite{ulayer} accelerates individual layers through cooperative execution and processor-friendly quantization.
MOSAIC~\cite{MOSAIC} slices and executes models while accounting for processor heterogeneity, communication costs, and execution constraints, while AsyMo~\cite{asymo} optimizes inference on asymmetric mobile CPUs.
At the multi-application level, NestDNN~\cite{nestdnn} dynamically selects resource--accuracy trade-offs for concurrently running vision models according to available resources.
LegoDNN~\cite{legodnn} enables finer-grained adaptation through block-level model scaling.
These approaches focus on DNN execution and model adaptation. More closely related, MCSched~\cite{mei2026mcsched} combines memory-controller pressure with foreground timing feedback to suspend and resume background LLM inference on Jetson. 
\system provides safe sublayer suspension across prefill and decode, monitors foreground memory demand and stalls, and schedules inference according to both foreground deadlines and LLM output freshness.

\vspace{0.05in}
\noindent\textbf{Inference preemption and accelerator scheduling.}
Existing systems control inference at different boundaries: FastServe preempts decoding between output tokens~\cite{wu2026fastserve}, Sarathi-Serve divides prefill into token chunks~\cite{agrawal2024sarathi}, and Pantheon partitions DNN graphs into preemptible chunks~\cite{han2024pantheon}. Laser and FlowPrefill provide control at Transformer-layer and operator boundaries, respectively~\cite{liao2026laser,hsieh2026flowprefill}. At the accelerator level, XSched exposes preemptible command queues across diverse accelerators~\cite{shen2025xsched}, while Bless combines fine-grained kernel scheduling with spatial resource allocation to preserve application GPU quotas~\cite{zhang2025bless}. Conspirator~\cite{xiao2025conspirator} coordinates data delivery and accelerator scheduling through a SmartNIC-based control plane for distributed ML. GPUWeaver~\cite{chen2026gpuweaver} and SwiftRDMA~\cite{swiftrdma} extend application-aware scheduling to shared GPU fabrics and RNIC resources, respectively. SERENO uses speculative decoding to add yield points to mobile LLM inference and release bandwidth to foreground applications~\cite{xin2026sereno}; its control mechanism is tied to the draft-and-verify workflow. 
\system provides suspension points across prefill and decode and schedules them for workflows in which foreground observations become LLM inputs and the resulting feedback has a freshness deadline. 

\vspace{0.05in}
\noindent\textbf{Execution monitoring.}
Existing tools expose complementary views of concurrent execution. 
Profilers such as ProfInfer~\cite{zou2026profinfer} and eInfer~\cite{chu2025einfer} uses eBPF events to reconstruct fine-grained or distributed inference behavior~\cite{zou2026profinfer,chu2025einfer}. Neutrino probes GPU execution at instruction granularity~\cite{huang2025neutrino}, and NVIDIA Nsight Systems correlates CPU, GPU, and hardware activity for performance diagnosis~\cite{nvidia2026nsightsystems}. Such detail is valuable for analysis but costly to collect and deliver continuously on the scheduling path. PMU counters offer a lighter source of resource-use information~\cite{nvidia2026pmu,qualcommPerformanceTools,linuxPerfEventOpen}. \system uses periodic foreground counter samples to derive memory-demand and stall signals, then publishes them for online scheduling without constructing full execution traces.


\section{Conclusion}


We presented \system{}, a bandwidth-aware runtime for co-running
LLM inference and latency-critical tasks on unified-memory edge devices.
\system combines safe sublayer suspension with lightweight foreground
monitoring to schedule inference according to contention risk, deadline
slack, and output freshness.
Across four edge platforms and two workloads, \system reduces foreground
deadline misses by 24.1--96.1\% relative to unmanaged inference while
retaining 64.9--98.5\% of its throughput, with corresponding improvements
in task quality.


\FloatBarrier
\bibliographystyle{ACM-Reference-Format}
\bibliography{references}

\clearpage
\appendix
\section{Complete Pause/Resume Measurements}
\label{sec:appendix-transitions}

This appendix expands the P99 summary in Table~\ref{tab:micro-transitions}
with all reported P50/P95/P99 latencies and sample counts.
Table~\ref{tab:appendix-hold} reports the delay from a \texttt{Hold}
decision to drain completion. Tables~\ref{tab:appendix-admit}
and~\ref{tab:appendix-resume} report resumption delays measured from
an \texttt{Admit} decision and the end of a \texttt{Hold} interval,
respectively, to the next subgraph submission. Both resumption
measurements exclude subsequent GPU queueing. All latencies are
in milliseconds; a dash marks an unreported sample count.

Across all reported phases, the delay from a \texttt{Hold} decision
to stopping further submissions and reaching the legal hold boundary
is recorded as 0\,ms. GPU service stops only after outstanding work
drains; its reported latency equals the drain latency in
Table~\ref{tab:appendix-hold}. Thus, zero submission delay does not
imply instantaneous GPU preemption.


\begin{table}[H]
    \centering
    \small
    \setlength{\tabcolsep}{5pt}
    \caption{\texttt{Hold} decision to drain completion (ms).}
    \label{tab:appendix-hold}
    \begin{tabular}{lrrrr}
        \toprule
        Phase & Samples & P50 & P95 & P99 \\
        \midrule
        \multicolumn{5}{l}{Orin Nano Super: Qwen2.5-1.5B} \\
        All & 109 & 0.000 & 3.840 & 4.176 \\
        Prefill & 68 & 0.111 & 2.931 & 3.989 \\
        Decode & 41 & 0.000 & 4.078 & 5.429 \\
        \midrule
        \multicolumn{5}{l}{AGX Orin: Qwen3-8B} \\
        All & -- & 1.025 & 2.081 & 3.202 \\
        Prefill & -- & 1.080 & 2.303 & 3.348 \\
        Decode & -- & 1.020 & 1.532 & 2.906 \\
        \midrule
        \multicolumn{5}{l}{AGX Orin: Qwen3-VL-8B} \\
        All & 11,020 & 1.246 & 2.873 & 3.719 \\
        Prefill & 6,261 & 1.341 & 2.352 & 3.371 \\
        Decode & 2,290 & 1.128 & 1.449 & 2.756 \\
        Vision & 2,469 & 2.468 & 3.109 & 4.724 \\
        \bottomrule
    \end{tabular}
\end{table}

\newpage
\begin{table}[H]
    \centering
    \small
    \setlength{\tabcolsep}{5pt}
    \caption{\texttt{Admit} decision to next subgraph submission (ms).}
    \label{tab:appendix-admit}
    \begin{tabular}{lrrrr}
        \toprule
        Phase & Samples & P50 & P95 & P99 \\
        \midrule
        \multicolumn{5}{l}{Orin Nano Super: Qwen2.5-1.5B} \\
        All & 542 & 0.297 & 0.881 & 2.689 \\
        Prefill & 358 & 0.271 & 0.768 & 1.865 \\
        Decode & 184 & 0.309 & 1.188 & 3.040 \\
        \midrule
        \multicolumn{5}{l}{AGX Orin: Qwen3-8B} \\
        All & 15,227 & 0.112 & 0.551 & 1.221 \\
        Prefill & 7,269 & 0.234 & 0.619 & 1.515 \\
        Decode & 7,958 & 0.093 & 0.487 & 0.847 \\
        \midrule
        \multicolumn{5}{l}{AGX Orin: Qwen3-VL-8B} \\
        All & 9,866 & 0.183 & 0.713 & 1.546 \\
        Prefill & 5,610 & 0.284 & 0.952 & 1.774 \\
        Decode & 2,123 & 0.088 & 0.387 & 0.694 \\
        Vision & 2,133 & 0.091 & 0.142 & 0.617 \\
        \bottomrule
    \end{tabular}
\end{table}

\begin{table}[H]
    \centering
    \small
    \setlength{\tabcolsep}{5pt}
    \caption{End of \texttt{Hold} to next subgraph submission (ms).}
    \label{tab:appendix-resume}
    \begin{tabular}{lrrrr}
        \toprule
        Phase & Samples & P50 & P95 & P99 \\
        \midrule
        \multicolumn{5}{l}{Orin Nano Super: Qwen2.5-1.5B} \\
        All & 74 & 0.468 & 1.675 & 3.037 \\
        Prefill & 43 & 0.514 & 0.666 & 2.229 \\
        Decode & 31 & 0.377 & 2.533 & 2.835 \\
        \midrule
        \multicolumn{5}{l}{AGX Orin: Qwen3-8B} \\
        All & 15,133 & 0.190 & 0.675 & 1.713 \\
        Prefill & 7,269 & 0.288 & 0.784 & 1.961 \\
        Decode & 7,864 & 0.157 & 0.586 & 1.329 \\
        \midrule
        \multicolumn{5}{l}{AGX Orin: Qwen3-VL-8B} \\
        All & 9,792 & 0.237 & 0.840 & 1.806 \\
        Prefill & 5,579 & 0.356 & 1.058 & 2.176 \\
        Decode & 2,084 & 0.153 & 0.488 & 1.125 \\
        Vision & 2,129 & 0.152 & 0.260 & 0.975 \\
        \bottomrule
    \end{tabular}
\end{table}

\twocolumn[{
\begin{minipage}{\textwidth}
\section{Case Study 1: Sensitivity}
\label{sec:appendix-case1}
\centering
\includegraphics[width=\textwidth]{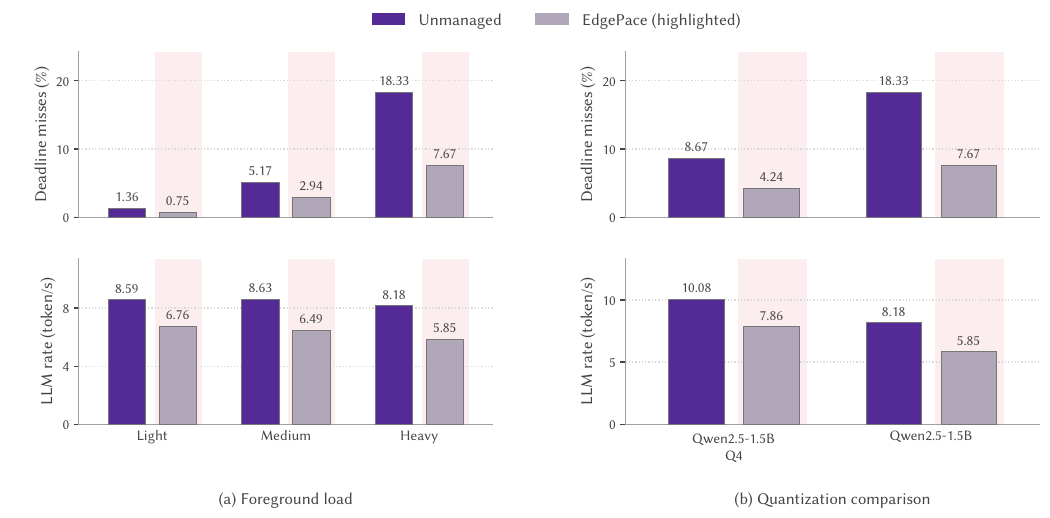}
\captionsetup{hypcap=false}
\captionof{figure}{Supplementary Qwen2.5-1.5B results on Orin Nano Super.
(a) Three foreground loads.
(b) The Q4 variant and default model with four camera inputs.
Upper plots show deadline-miss rates; lower plots show LLM token rates.
Light red bands highlight \system. The heavy-load and default-model points
use the four-camera results from Figure~\ref{fig:case1-performance};
light/medium-load and Q4 points retain the earlier measurements.}
\label{fig:case1-sensitivity}
\Description{Four paired bar charts compare Unmanaged and EdgePace with
Qwen2.5-1.5B across light, medium, and heavy foreground loads, and for the
Q4 variant and default model. EdgePace reduces misses while retaining part
of the unmanaged token throughput.}
\end{minipage}
\vspace{0.10in}
}]

\noindent\textbf{Sensitivity to foreground load.}
Across light, medium, and heavy loads, \system reduces misses from
1.36\%, 5.17\%, and 18.33\% to 0.75\%, 2.94\%, and 7.667\%,
respectively. Throughput retention is 78.7\%, 75.2\%, and 71.5\%
(Figure~\ref{fig:case1-sensitivity}(a)). In the updated four-camera run,
misses fall by 58.2\%. Throughput retention remains below that of the
earlier runs at lighter loads, illustrating the cost of protecting a
heavier foreground workload.

\vspace{0.05in}
\noindent\textbf{Sensitivity to quantization.}
Figure~\ref{fig:case1-sensitivity}(b) compares the Q4 variant with the
default Qwen2.5-1.5B model. For Q4, \system reduces misses from
8.67\% to 4.24\% (51.1\% fewer), while throughput falls from
10.08 to 7.86\,token/s (78.0\% retained). For the default model, misses fall
from 18.33\% to 7.667\% (58.2\% fewer), while throughput decreases
from 8.175 to 5.848\,token/s (71.5\% retained). Both configurations
benefit from scheduling, with a larger throughput trade-off for the
default model in these runs.

\twocolumn[{
\begin{minipage}{\textwidth}
\section{Scheduling Decision Distributions}
\label{sec:appendix-risk}

Figure~\ref{fig:appendix-risk} and Table~\ref{tab:appendix-risk} report
\system's decisions across the 12 recorded case-study configurations.
The log's \texttt{normal} category denotes \textsc{Bypass}. Percentages
count scheduler checks, including repeated checks during \textsc{Hold};
\textsc{Admit} includes decisions caused by freshness urgency as well as
medium contention risk (\cref{sec:scheduler}). Hold time is the cumulative
time spent waiting in \textsc{Hold} during each run.

\medskip
\centering
\includegraphics[width=\textwidth]{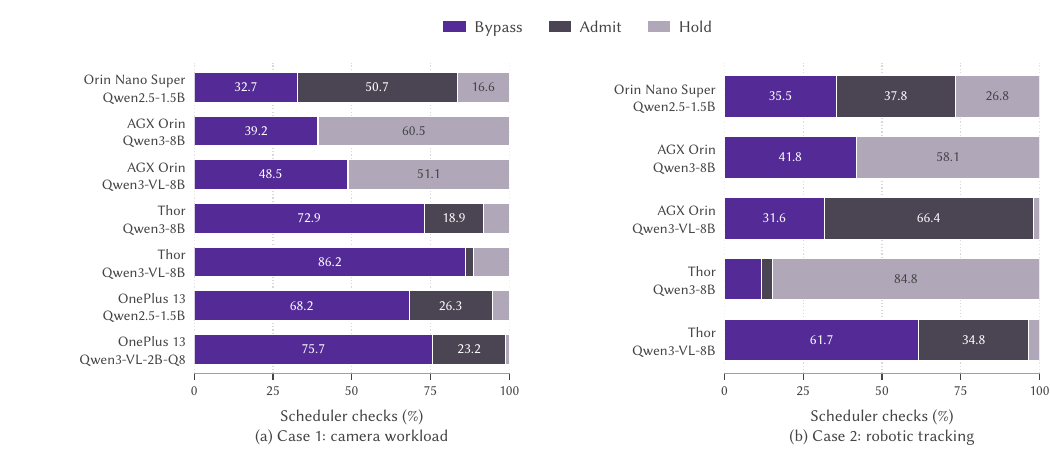}
\captionsetup{hypcap=false}
\captionof{figure}{Scheduler decision distributions for (a) the camera
workload and (b) robotic target tracking. Each bar represents one
platform--model configuration running \system. Segments show the
percentage of checks selecting each mode; labels are shown for segments
of at least 12\%. Table~\ref{tab:appendix-risk} provides all percentages
and cumulative hold times.}
\label{fig:appendix-risk}
\Description{Stacked horizontal bars show Bypass, Admit, and Hold shares
for seven camera and five robotic-tracking configurations. Hold shares
range from 1.13 to 60.52 percent for the camera workload and from 2.00
to 84.82 percent for robotic tracking. Percentages count scheduler
checks rather than elapsed time.}

\vspace{0.08in}
\captionof{table}{Complete decision distributions and cumulative hold
times. Percentages and times are rounded independently to two and three
decimal places, respectively; percentage totals may differ slightly
from 100 due to rounding.}
\label{tab:appendix-risk}
{\small\setlength{\tabcolsep}{5pt}
\begin{tabular}{@{}lllrrrr@{}}
\toprule
Case & Platform & Model & Bypass (\%) & Admit (\%) & Hold (\%) & Hold time (s) \\
\midrule
1 & Orin Nano Super & Qwen2.5-1.5B & 32.71 & 50.66 & 16.63 & 18.815 \\
\addlinespace[2pt]
1 & AGX Orin & Qwen3-8B & 39.21 & 0.26 & 60.52 & 18.062 \\
1 & AGX Orin & Qwen3-VL-8B & 48.55 & 0.34 & 51.10 & 15.460 \\
\addlinespace[2pt]
1 & Thor & Qwen3-8B & 72.93 & 18.93 & 8.15 & 19.553 \\
1 & Thor & Qwen3-VL-8B & 86.17 & 2.39 & 11.44 & 7.545 \\
\addlinespace[2pt]
1 & OnePlus 13 & Qwen2.5-1.5B & 68.22 & 26.26 & 5.52 & 1.611 \\
1 & OnePlus 13 & Qwen3-VL-2B-Q8 & 75.69 & 23.18 & 1.13 & 0.435 \\
\midrule
2 & Orin Nano Super & Qwen2.5-1.5B & 35.46 & 37.76 & 26.78 & 3.432 \\
\addlinespace[2pt]
2 & AGX Orin & Qwen3-8B & 41.76 & 0.15 & 58.09 & 5.878 \\
2 & AGX Orin & Qwen3-VL-8B & 31.57 & 66.43 & 2.00 & 2.960 \\
\addlinespace[2pt]
2 & Thor & Qwen3-8B & 11.73 & 3.46 & 84.82 & 6.288 \\
2 & Thor & Qwen3-VL-8B & 61.69 & 34.83 & 3.49 & 1.467 \\
\bottomrule
\end{tabular}
}
\end{minipage}
\vspace{0.10in}
}]

\noindent\textbf{Variation across configurations.}
The selected modes vary substantially with the workload and model.
In Case~1, \textsc{Hold} accounts for 1.13--60.52\% of checks, while
Thor selects \textsc{Bypass} for 72.93\% and 86.17\% of checks with its
text and vision-language models. In Case~2, AGX Orin's Qwen3-8B run
selects \textsc{Hold} for 58.09\% of checks, whereas its Qwen3-VL-8B
run predominantly selects \textsc{Admit} (66.43\%). These are
distributions of selected actions after freshness overrides.

\vspace{0.05in}
\noindent\textbf{Decision frequency and waiting time.}
Thor's Qwen3-8B run selects \textsc{Hold} for 84.82\% of checks in
Case~2 but accumulates 6.288\,s of waiting, compared with 8.15\% and
19.553\,s in Case~1. The scheduler repeats checks while held, so a
large \textsc{Hold} share does not imply a correspondingly large
share of elapsed time. Run durations are not recorded with these
measurements, so we report cumulative waiting times without converting
them to runtime fractions.

\twocolumn[{
\begin{minipage}{\textwidth}
\section{Graph Partitioning: Phase Breakdown}
\label{sec:appendix-partition}

Table~\ref{tab:appendix-partition} expands the totals in
Table~\ref{tab:micro-partition} with separate vision, prefill,
and decode timings for all seven platform--model configurations.

\medskip
    \centering
    \small
    \renewcommand{\arraystretch}{1.10}
    \captionsetup{hypcap=false}
    \setlength{\tabcolsep}{4pt}
    \captionof{table}{Execution time with and without graph partitioning
    for 512 prefill tokens and 128 decode tokens; VLMs use a
    512-token visual budget. Vision denotes vision encoding,
    and VLM prefill excludes that stage. The final column is
    the percentage increase in total time, calculated before
    rounding. Reported totals may differ from sums of rounded
    phase timings. A dash denotes an inapplicable stage.}
    \label{tab:appendix-partition}
    \begin{tabular}{@{}lrrrrrrrrr@{}}
        \toprule
        {\sffamily\bfseries Model} &
        \multicolumn{4}{c}{\sffamily\bfseries Unpartitioned (s)} &
        \multicolumn{4}{c}{\sffamily\bfseries Partitioned (s)} &
        {\sffamily\bfseries Increase} \\
        \cmidrule(lr){2-5}
        \cmidrule(lr){6-9}
        & {\sffamily\bfseries Vision} & {\sffamily\bfseries Prefill} & {\sffamily\bfseries Decode} & {\sffamily\bfseries Total}
        & {\sffamily\bfseries Vision} & {\sffamily\bfseries Prefill} & {\sffamily\bfseries Decode} & {\sffamily\bfseries Total} & {\sffamily\bfseries (\%)} \\
        \midrule
        \multicolumn{10}{l}{\sffamily\bfseries Orin Nano Super} \\
        Qwen2.5-1.5B
            & --- & 0.34 & 7.05 & 7.39
            & --- & 0.45 & 7.25 & 7.70 & 4.19 \\
        \midrule
        \multicolumn{10}{l}{\sffamily\bfseries AGX Orin} \\
        Qwen3-8B
            & --- & 0.624 & 5.120 & 5.744
            & --- & 0.630 & 5.290 & 5.920 & 3.07 \\
        Qwen3-VL-8B
            & 0.609 & 1.592 & 5.351 & 7.553
            & 0.614 & 1.613 & 5.345 & 7.572 & 0.25 \\
        \midrule
        \multicolumn{10}{l}{\sffamily\bfseries Jetson Thor} \\
        Qwen3-8B
            & --- & 0.551 & 3.273 & 3.824
            & --- & 0.550 & 3.292 & 3.842 & 0.46 \\
        Qwen3-VL-8B
            & 0.361 & 1.113 & 3.347 & 4.821
            & 0.338 & 1.123 & 3.387 & 4.849 & 0.57 \\
        \midrule
        \multicolumn{10}{l}{\sffamily\bfseries OnePlus 13} \\
        Qwen2.5-1.5B
            & --- & 0.353 & 7.716 & 8.069
            & --- & 0.358 & 8.074 & 8.432 & 4.50 \\
        Qwen3-VL-2B-Q8
            & 8.855 & 2.071 & 15.102 & 26.027
            & 8.861 & 2.068 & 15.745 & 26.673 & 2.48 \\
        \bottomrule
    \end{tabular}


\end{minipage}
\vspace{0.10in}
}]

\noindent\textbf{Variation across phases.}
On Orin Nano Super, prefill time increases by 32.4\%, versus
2.8\% for decode. For the OnePlus 13 VLM, decode time increases
by 4.26\%, while vision encoding and prefill change by less
than 0.2\% each, yielding a smaller total increase of 2.48\%.

\end{document}